\documentclass[]{tMPH2e}

\def\cpl{Chem. Phys. Lett.~}

\def\pra{{Phys.~Rev.~A~}}

\def\prl{{Phys.~Rev.~Lett.~}}

\def\jcp{{J.~Chem.~Phys.~}}

\usepackage{color}
\usepackage[frenchb,english]{babel}
\usepackage{amsmath}
\usepackage{dcolumn}
\usepackage{graphicx}
\usepackage{dcolumn}
\usepackage{bm}
\usepackage{rotating}
\usepackage{amssymb}
\usepackage[latin1,utf8]{inputenc}
\usepackage{natbib}
\usepackage[Symbol]{upgreek}
\usepackage{epstopdf}
\graphicspath{{./figures/}}

\begin{document}
\doi{10.1080/0026897YYxxxxxxxx}
\issn{1362-3028}
\issnp{0026-8976}
\jvol{00}
\jnum{00} \jyear{2012} 

\markboth{I. Manai et al}{Molecular Physics}

\articletype{INVITED ARTICLE}
\title{\itshape {Laser cooling of rotation and vibration by optical pumping}}
\author{I. Manai$^{a\ast}$, R. Horchani$^{a}$, M. Hamamda$^{a}$, A. Fioretti$^{b}$, M. Allegrini$^{b}$, H. Lignier$^{a}$, P. Pillet$^{a}$\thanks{$^\ast$Corresponding author. Email: isam.manai@u-psud.fr
\vspace{6pt}} and D. Comparat$^{a}$\\\vspace{6pt}  $^{a}${\em{Laboratoire Aim\'e Cotton, CNRS, Universit\'e Paris-Sud 11, ENS Cachan, B\^at 505, Campus d'Orsay, 91405 Orsay, France}};
$^{b}${\em{Dipartimento di Fisica, Universit\`a di Pisa and INO-CNR Sezione di Pisa, Largo Pontecorvo 3, 56127 Pisa, Italy}}\\\vspace{6pt}\received{v0 released December 2012}}

\maketitle
\date{\today}
\begin{abstract}
We have recently demonstrated that optical pumping methods combined with photo-association of ultra-cold atoms can produce ultra-cold and dense samples of molecules in their absolute rovibronic ground state. More generally, both the external and internal degrees of freedom can be cooled by addressing selected rovibrational levels on demand. Here, we recall the basic concepts and main steps of our experiments, including the excitation schemes and detection techniques we use to achieve the rovibrational cooling of $\rm Cs_{2}$ molecules. In addition, we present the determination of formation pathways and a theoretical analysis explaining the experimental observations. These simulations improves the spectroscopic knowledge required to transfer molecules to any desired rovibrational level.
\end{abstract}
\bigskip

\bigskip

\section{Introduction}
The precise control of both the internal and external degrees of freedom of cold and ultra-cold molecules is very attractive and promises new insights and advances in various domains of physics \cite{LincolnDCarr09, 2011_Nat_Hudson, 2012_JinYe_Introduction}  and modern physical chemistry \cite{Shapiro03, Alessandro2007, 2012_ChemRev_Quemener_MolControl}. Several theoretical approaches have been proposed to control the internal degrees of freedom of cold molecules, such as the use of an external cavity to favor spontaneous emission toward the lowest ro-vibrational level \cite{2007_PRL_Cavity_Cooling} or the controlled interplay of coherent laser fields and spontaneous emission through quantum interferences between different transitions \cite{1999FaraDisc, bartana1993, 2001PhRvA..63a3407S, 2009Brif_QuantumContro}. From the experimental point of view, molecules are more difficult to control than atoms because of their complex internal structure. However, cold molecule formation has been achieved by different approaches that can be roughly grouped into two main classes : direct cooling of preformed molecules and indirect cold molecule formation by assembling cold or ultra-cold atoms \cite{LincolnDCarr09}. In summary, two major challenges have been faced, the first being the decrease of the translational temperature of molecules in analogy with atom cooling and the second being the reduction of the internal rovibronic energy. Ultimately, experimental efforts have been focused on the control of the external degrees of freedom, leading to impressive achievements :  one-dimensional transverse laser cooling \cite{shuman2010laser}, longitudinal slowing \cite{PhysRevLett.108.103002} of a SrF beam, and more recently, one- and two-dimensional laser cooling with magneto-optical trapping of polar YO molecules \cite{2012YeMolTrapping}, Sisyphus cooling of electrically trapped polyatomic CH$_{3}$F molecules \cite{2012_SysCoolingMol} and evaporative cooling of molecules \cite{2012_EvapCoolingMol}. Other techniques have been also developed to coherently transfer molecular populations from an hyperfine level to another one (Cs$_{2}$ \cite{danzl2010ultracold}, KRb \cite{2010PhRvL.104c0402O}).

In our laboratory, we have been able to produce Cs$_{2}$ molecules with low translational energy (T $\simeq 100 \ \mu K $) by photo-association of ultra-cold Cs atoms as soon as 1998 \cite{Fio1}. Unfortunately, those molecules have a substantial residual internal energy because the decay of the electronically excited state used in the photo-association process generally leads to populate several vibrational and rotational levels. We have only recently achieved the transfer and the accumulation of molecules to a single rovibrational level of the electronic ground state \cite{MatthieuViteau07112008, PhysRevLett.109.183001}. The aim of this paper is to describe the basics of our method that is, in summary, based on a broadband and incoherent optical source designed to modify the distribution of the ro-vibrational populations. Our approach is actually a follow up of the so called "luminorefrigeration" process proposed by Kastler \cite{kastler1950} in 1950 to accumulate sodium atoms in the lowest state of the hyperfine manifold. Later on, it was also proposed to lower the "J" value of Cs$_{2}$ molecules \cite{1996JChPh.104.9689B}. The principle is quite simple and general and has been successfully applied to decrease the vibrational temperature of NaCs \cite{2012_OptExpr_Bigelow_NaCs_Cooling} and cool molecular ions \cite{2010NatPhDrewsen_rotational_cooling, 2010NatPh...6..275S}. In all these works,  molecules present in several levels are optically excited and decay to other levels as long as they reach a target level which is not accessible ("dark") to the optical excitation. Such an attainment has required great efforts to find out appropriate schemes and consequent technical hints for both the excitation and detection. For the vibrational cooling of Cs$_{2}$ molecules, these details have been already reported. On the other hand, rotational cooling of Cs$_{2}$ is a more subtle process given that the molecular vibration is likely to be affected. Finding a solution for simultaneous rotational and vibrational cooling has been achieved and it is described in the following sections. 
This article is organized in the following way. In section 2, we explain the general photo-association process producing bound molecules and we describe our detection methods that is crucial for the diagnostics. In section 3, the vibrational cooling of translational cold Cs$_{2}$ molecules obtained by an improved laser shaping is presented. In section 4, we describe spectroscopic investigations of the Cs$_{2}$ rotational structure and simulations relevant for the rotational cooling. In section 5, we report the details of the simultaneous vibrational and rotational cooling by optical pumping methods.

\section{Photoassociation}
Photoassociation (PA) of atoms in atomic vapors is a well-known process \cite{1983Pichler_JPhysB, Jones93, Marvet95, Ban99}. In the specific case of cesium, a pair of atoms in hyperfine level $F$ of the $6s$ ground state absorbs one photon of suitable energy h$\rm{\nu_{PA}}$, red-detuned from the energy of the atomic transition ($6s + 6p_j$, $j = \frac{1}{2}$ or $ \frac{3}{2}$) to create molecules in a rovibrational level ($v$,$J$) of an excited electronic state $\Omega$ related to one of the asymptotic ($6s + 6p_{1/2}$) or ($6s + 6p_{3/2}$) according to the reaction

\begin{eqnarray*}
 {\rm Cs}(6s,F) + {\rm Cs}(6s,F) + h\nu_{\rm PA} \rightarrow {\rm Cs}_2^* (\Omega(6s + 6p_j);v,J)
\end{eqnarray*}

For atoms at room temperature, the PA process does not allow to resolve the excited state level structure because the width of the Maxwell-Boltzmann kinetic energy distribution of the atoms, $k_BT$ ($k_B$ is the Boltzman constant and $T$ the temperature of the atomic sample), is much larger than the energy spacing in the excited state \cite{LeonidRybak11}. On the contrary, due to the extremely narrow width of the thermal distribution of ultracold atoms ($k_BT \approx h\times 2$ MHz at $ T \approx 100$ $\umu$K), smaller than the energy spacing between molecular bound levels, and more importantly, very often narrower than the level width of the excited states, photo-association of cold atoms has proven to be a powerful tool for high-resolution molecular spectroscopy \cite{Jones06}. It has given access to the previously unexplored domain of molecular dynamics at distances well beyond those of well-known chemical bonds. Indeed, the photo-association process occurs at large inter-atomic distances $R$, i.e. roughly at the classical outer turning point of the final vibrational level, where a pair of identical ground state atoms interact through their Van der Waals interaction behaving as $R^{-6}$. When the atom pair is excited, the dipole-dipole interaction is dominant, and varies as $R^{-3}$. Vibrational levels with a very large elongation (beyond $100$ a$_0$) are then efficiently populated by photo-association. It is useful to characterize the PA process by the rate of excitation per atom in the sample, R$_{PA}$, which is generally proportional to ${\rm I}_L{\rm q}\rho T$, where ${\rm I}_{L}$ is the PA laser irradiance, $q$ is the Franck-Condon factor describing the overlap between the initial free state and the final bound state, $\rho$ is the phase space density \cite{PRABohn99, 2000IJQE...36.1378D}. Depending on the shape of the potential curve of the excited state, the excited molecules decay, with typical times of the order of tens of nanoseconds, by a spontaneous emission into two free "hot" atoms with a large relative kinetic energy, or into rovibrational levels of the lowest stable electronic states, which is the case of interest for the production of translationally cold molecules.

\subsection{Molecules in the singlet ground electronic state X$^1\Sigma_g^+$}
\begin{figure}
\begin{center}
\begin{minipage}{\textwidth}
\subfigure{
\resizebox*{6cm}{!}{\includegraphics{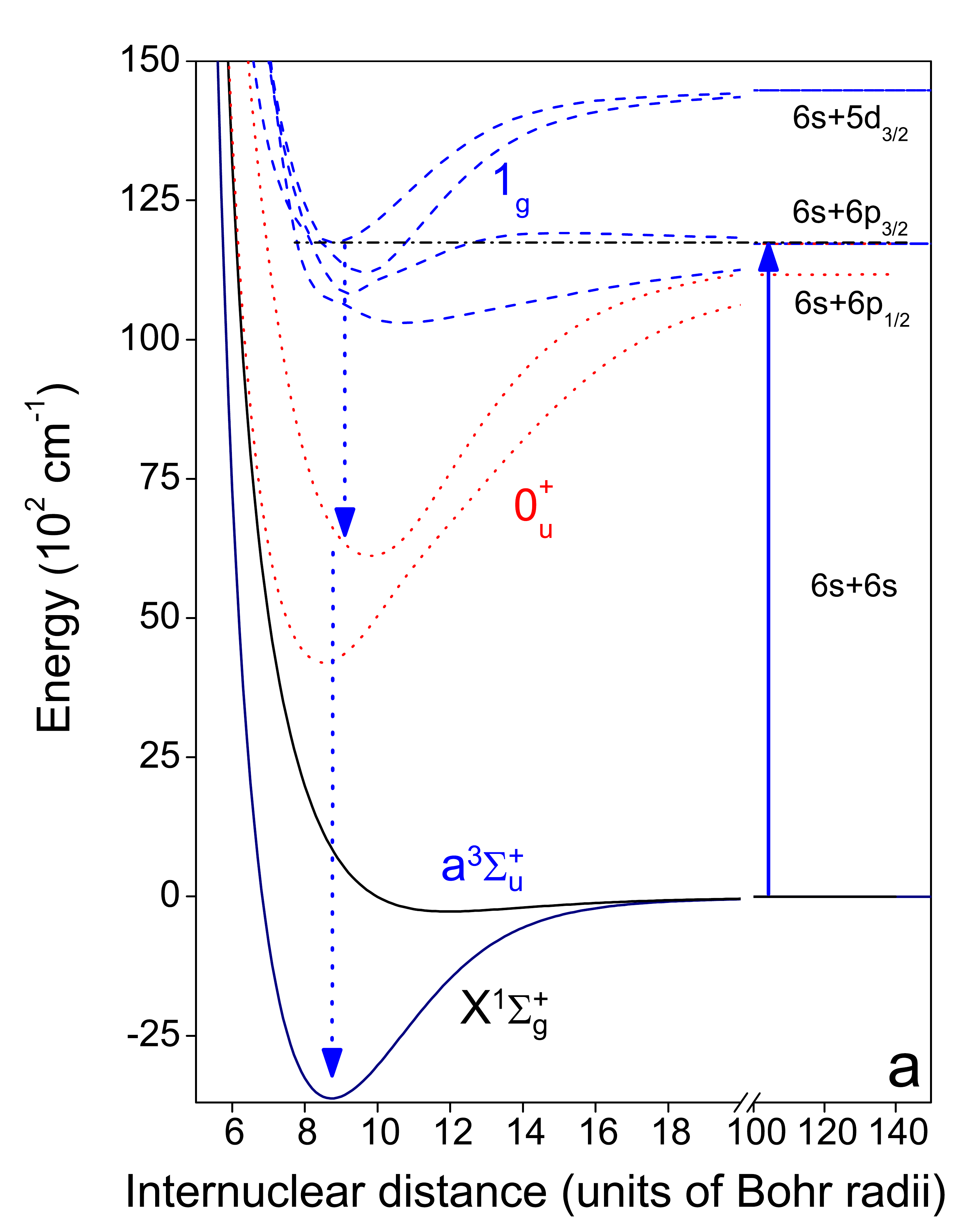}}}
\subfigure{
\resizebox*{6cm}{!}{\includegraphics{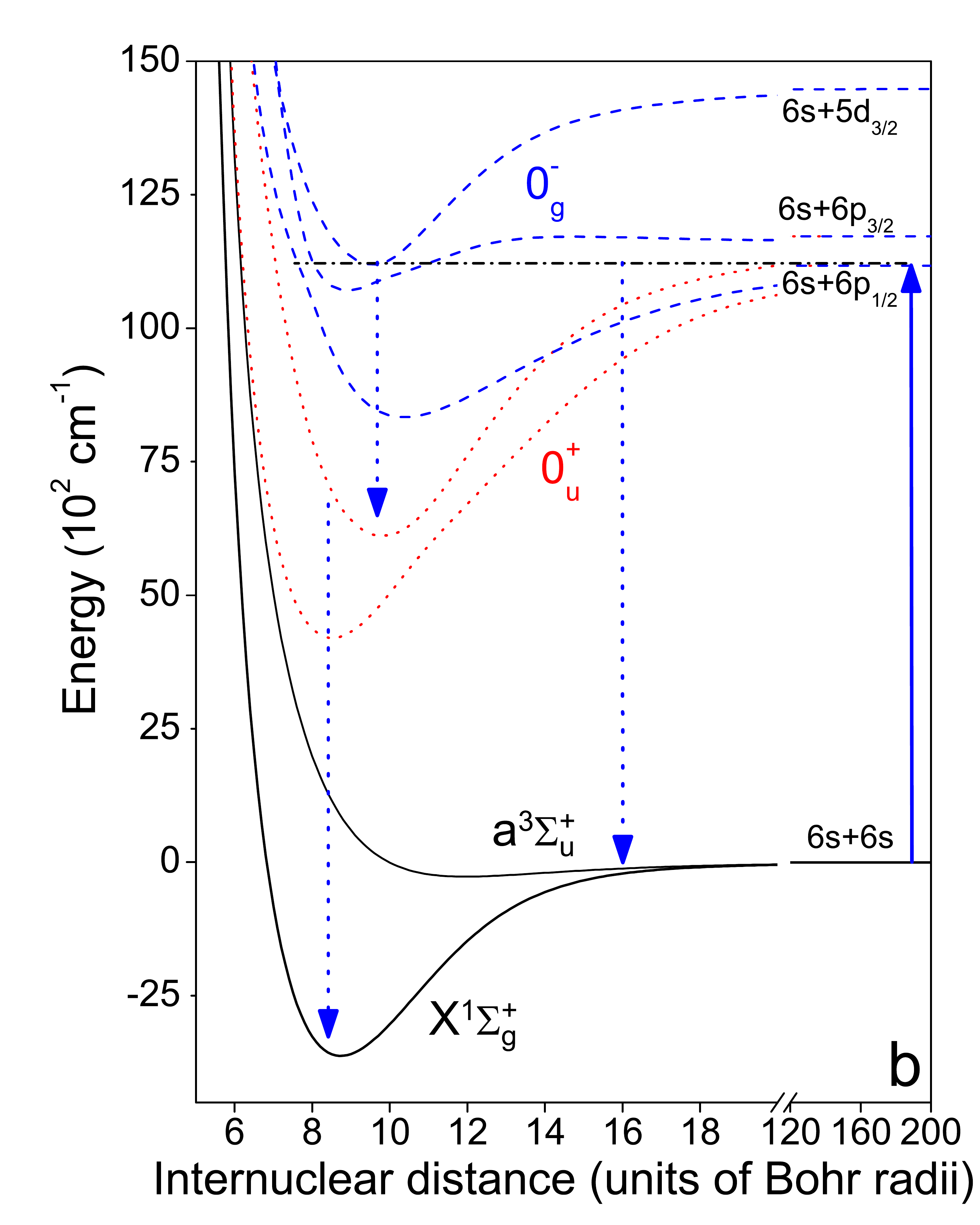}}}
\caption{Two photo-association schemes used to produce Cs$_{2}$ cold molecules in the X singlet ground electronic state. a) and b) correspond to photoassociation scheme through $1_g(6s+6p_{3/2})$ and $0_g^-(6s + 6p_{1/2})$ state, respectively. The solid line arrows correspond to the laser excitation, the dotted line arrows represent spontaneous emission paths.}
\label{PA}
\end{minipage}
\end{center}
\end{figure}

The core of our experiments is a Cs vapor-loaded MOT, where $10^{7}$ atoms are continuously laser-cooled and trapped at a temperature of T$\sim 100$ $\umu$K and at a peak density of $n \sim 10^{11}$ cm$^{-3}$. PA is achieved using a cw Titanium:Sapphire laser (intensity $\sim 300$ W.cm$^{-2}$) pumped by an Ar$^{+}$ laser. The PA scheme used to prepare cold cesium molecules in the X$^1\Sigma_g^+$ state (singlet ground electronic state hereafter referred to as X), schematically shown in Figure \ref{PA}a) was verified by a spectroscopic study using a broadband laser radiation as described in detail in \cite{SOFITIKIS}. The PA laser excites the atom pair to the ($v$, $J=8$) rovibrational level at 11730.1245 cm$^{-1}$ about 2 cm$^{-1}$ below the $6s+ 6p_{3/2}$ asymptote of the lowest $1_g(6s+6p_{3/2})$ long-range molecular potential. At short distances, this state is coupled to the lowest $1_g(6s+5d_{5/2}$) potential curve through several avoided crossings induced by spin-orbit interaction. The $v=0$ level of this curve is predicted to be very close to the $(6s+6p_{3/2})$ dissociation energy. It is surely the only populated short-range level in this process which allows for a decay to the X ground state through a two-photon spontaneous emission cascade via the $0_{u}^+(6s + 6p_{1/2})$ potentials, and finally producing nearly $10^6$ molecules per second. On the contrary, one-photon spontaneous emission favors the decay to levels of the triplet a$^{3}\Sigma_{u}^{+}$ state as reported in the first observation of the formation of translationally cold molecules \cite{Fio1}. The triplet ground electronic state, however, at the average distance $8.73$ a$_0$ corresponds to the range of the repulsive wall so that only dissociating pairs could be formed. More recently, another mechanism of formation of cesium molecules in the ground singlet electronic state has been found \cite{2011PCCP...1318910L}. In this mechanism, shown in Figure\ref{PA}b), the PA laser is set to excite long range vibrational levels belonging to the $0_g^-(6s + 6p_{1/2})$ state. In the first $20$ ${\rm cm^{-1}}$ below the atomic asymptote, vibrational levels are coupled to some inner potential curves presumably belonging to the upper $0_g^-(6s + 6p_{3/2})$ and $0_g^-(6s + 5d_{3/2})$ states. Since the $0_g^-$  is a state of gerade symmetry, any one-photon spontaneous decay ends up either into the ${\rm a}^{3}\Sigma^{+}_{u}$ ground state or into the $\Omega = 0_u^{+/-}$, $1_u$ and $2_u$ components of the ${\rm A}^{1}\Sigma^{+}_{u}$ or b$^3\Pi_u$ state located at intermediate energies. The 2$_u$ state should be metastable, nevertheless, a resonant coupling with the $0_u^+$(A$^1\Sigma_u^+$) state allows for a decay of the $0_u^+$(b$^3\Pi_u$) component into the X ground state. 

\subsection{Selective ionization spectroscopy}
\begin{figure}
\begin{center}
\begin{minipage}{10cm}
\centering
\resizebox{8cm}{!}{\includegraphics{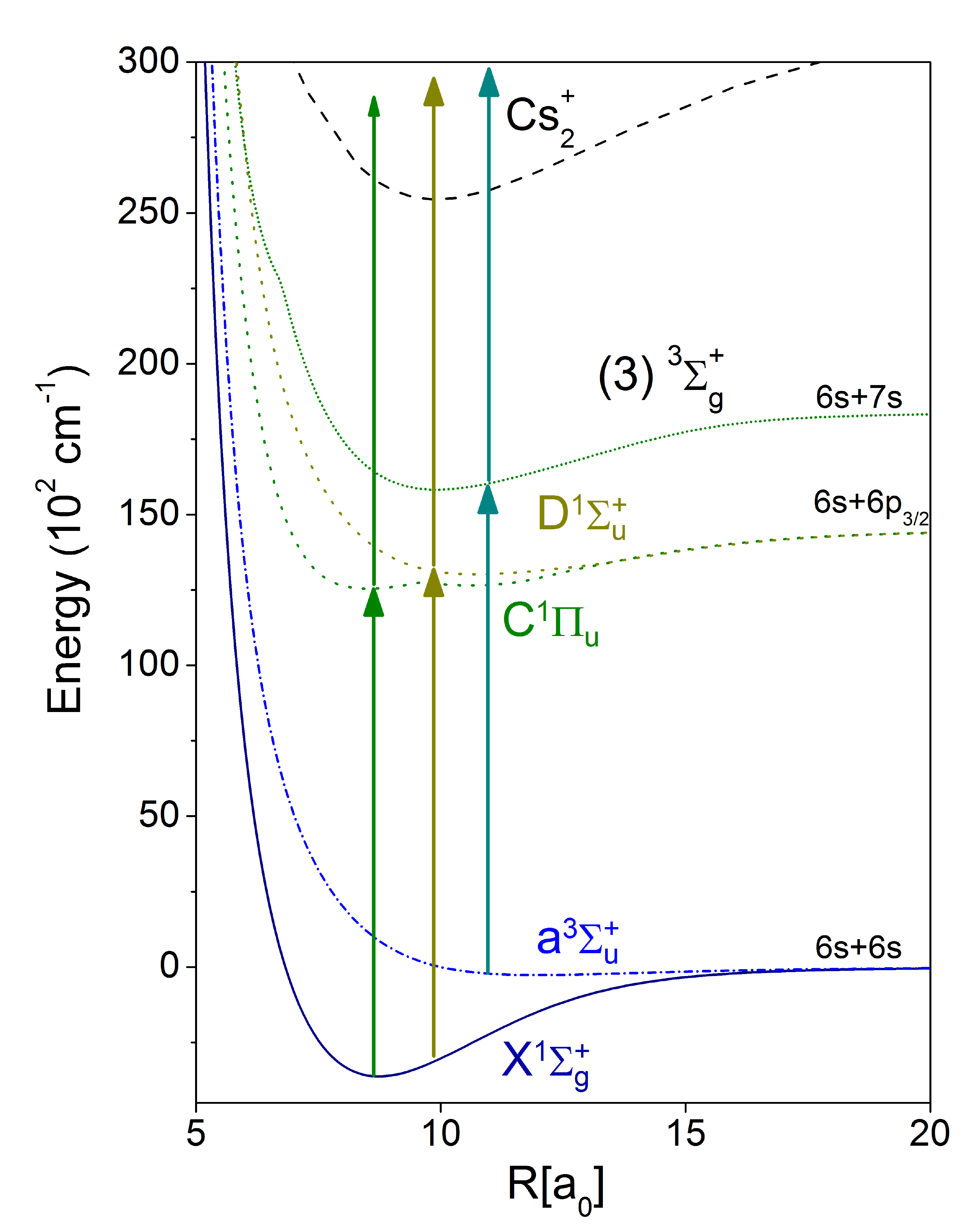}}
\caption{Cold molecule detection scheme by two photons ionization (see arrows). The C and D states are used to detect deeply bound molecules in the singlet X state while the $(3)^3\Sigma_g^+$ state is used to detect molecules in the triplet $a^{3}\Sigma_{u}^{+}$ state.}
\label{detection1}

\end{minipage}
\end{center}
\end{figure}

Whereas the Cs$_{2}$ molecules undergo a ballistic expansion and fall under the influence of gravity, a detection can be efficiently performed by Resonance Enhanced Multi-Photon Ionization (REMPI). This method relies on a selective production of molecular ions that are quickly extracted from the MOT region by a static electric field and detected by a pair of micro-channel plates (MCPs) through a time-of-flight mass spectrometer. As the ionization is based on resonant excitations, it is possible to get spectra by scanning the wavelength of the exciting pulsed dye laser pumped by the second harmonic of a pulsed Nd:YAG laser (repetition rate $10$ Hz, duration $7$ ns). 
In the case of Cs$_2$, the  spectral bandwidth ($\simeq 0.5$ ${\rm cm^{-1}}$) of the pulsed dye laser limits the resolution of the spectra to the vibrational structure. Electronic states generally involved in the REMPI process are schematically shown in Figure \ref{detection1}: in the $15700-16100$ ${\rm cm}^{-1}$ energy range, the ionization of deeply bound molecules in the X state proceeds via the C$^1\Pi_u(6s + 5d)$ state and the D$^1\Sigma_u^+ (6s + 5d)$ state. In figure \ref{PIspectrum} the assignment of the observed ionization lines is done according to the known spectroscopy \cite{1982JChPh..76.4370R, 1985JChPh..82.5354W, Yokoyama88, kato91}. In brief, this method, fully satisfying for vibrational analysis, is not sufficient to get information about the rotational distribution. Further necessary refinement in this perspective will be discussed in Section 4.

\section{Manipulation of the vibration with a broadband laser}
\subsection{Optical pumping}

\begin{figure}
\begin{center}
\begin{minipage}{100mm}
\subfigure{
\resizebox*{9cm}{!}{\includegraphics{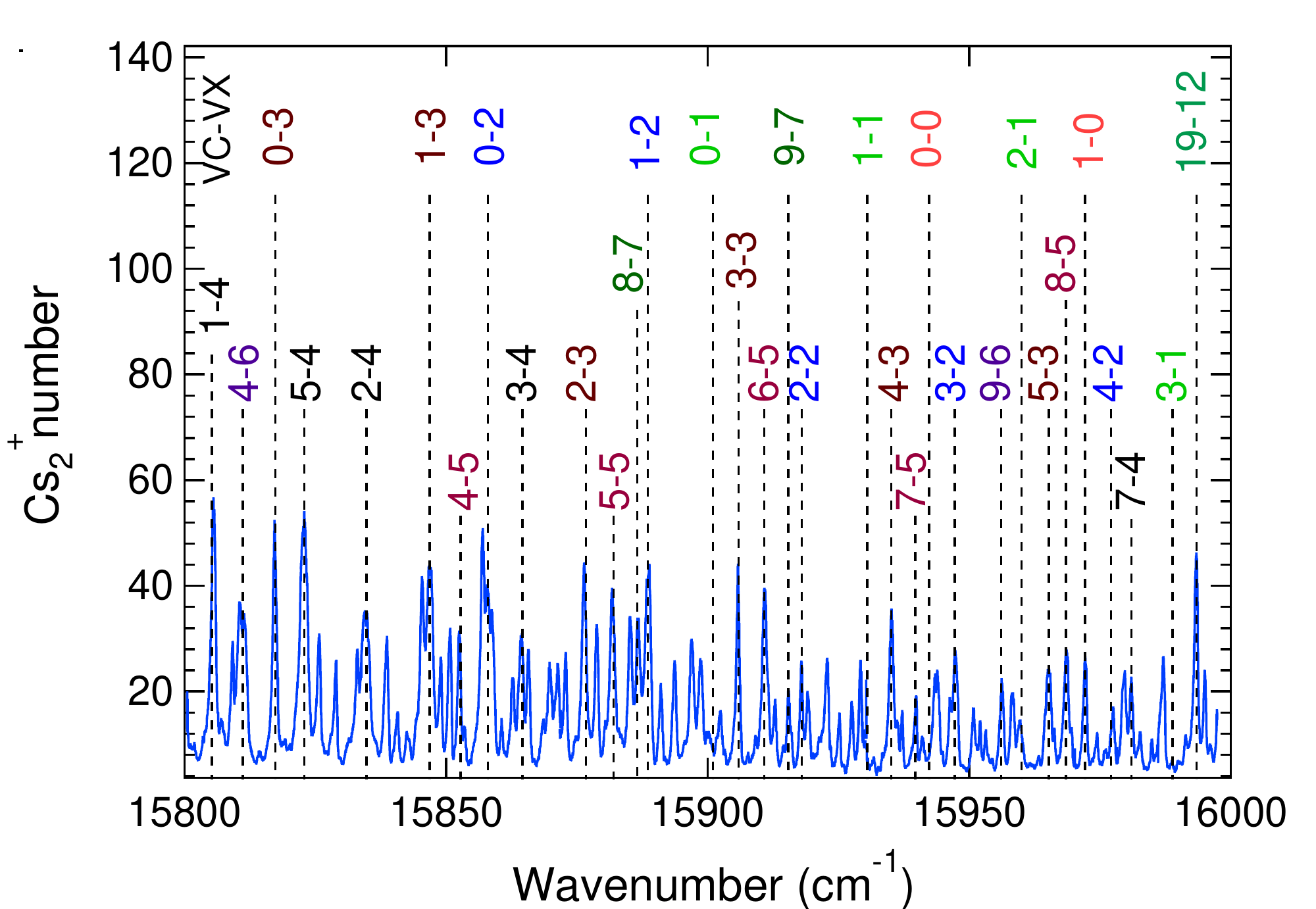}}}
\caption{(color online) Vibrational spectrum obtained by a 10 pulses average REMPI detection, corresponding to a PA process through the $0_g^{-}(\rm 6s + 6p_{1/2})$ potential. The pairs of numbers indicate the assignment of the $v_{C}\leftarrow v_{X}$ transitions resulting from a comparison with the spectroscopy of the C and X states.}
\label{PIspectrum}
\end{minipage}
\end{center}
\end{figure}

Experiments show that cold photo-associated molecules produced in the ground singlet X state are distributed over low vibrational levels $v_X = 0-10$ and relatively high rotational levels $J_X = 6-10$ when associated through the $1_g(6s+6p_{3/2})$ potential (see Figure \ref{PA}a.) while they are distributed over high vibrational levels $ v_X = 0-60$ and low rotational levels $J_X= 0-6$  when associated through the $0_g^{-}$($\rm 6s + 6p_{1/2}$) (shown in Figure \ref{PA}b.) potential. One example of such photo-ionization spectrum is shown in figure \ref{PIspectrum}. Starting from this observation, one of our first goal was to transfer these molecules in a well-defined vibrational level and, if possible, in the ground vibrational one.

The main idea of the vibrational pumping demonstrated for Cs$_2$ molecules in \cite{MatthieuViteau07112008, 2012_PRA_Horchani_Conversion, 2009NJP, Fioretti09, 2011PCCP...1318910L, PhysRevLett.109.183001} and recently for NaCs molecules \cite{2012_OptExpr_Bigelow_NaCs_Cooling} consists in using a broadband laser tuned to the transitions between the different vibrational levels $v_{X}$ and the levels belonging to an electronically excited state, namely the B$^{1}\Pi_{u}$ state in the cesium case. Starting from a given vibrational distribution of $v_{X}$, the aim is to transfer it into a single target vibrational level. We have first demonstrated this transfer of population from the vibrational levels of cold singlet-ground-state Cs$ _2$ molecules, towards the level, $v_{X}$ = $0$, with no vibration \cite{MatthieuViteau07112008}. Each absorption-spontaneous emission cycle lead, through optical pumping, to a new distribution of the vibrational population in the different vibrational levels of the ground state according to

\begin{eqnarray*}
\begin{aligned}
& {\rm Cs}_2(v_{\rm X}) +  h\nu_{\rm laser}  \rightarrow  {\rm Cs}_2^*(v_{\rm B})\ (\rm excitation) \\
& {\rm Cs}_2^*(v_{\rm B})  \rightarrow  {\rm Cs}_2(v'_{\rm X}) + h\nu_{\rm em}\ ({\rm spontaneous\ emission})
\end{aligned}
\end{eqnarray*}

\begin{figure}
\begin{center}
\begin{minipage}{100mm}
\resizebox{9cm}{!}{\includegraphics{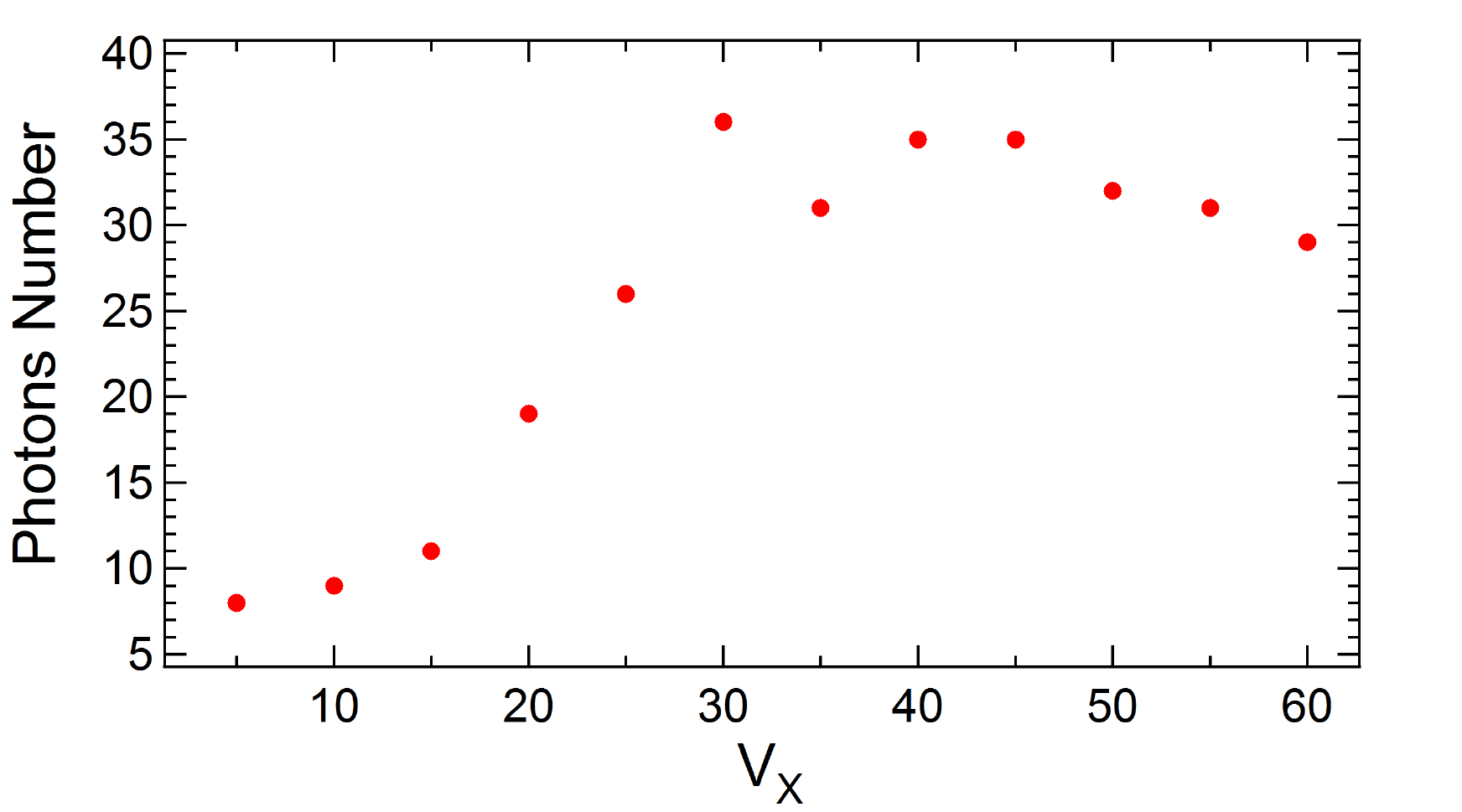}}
\caption{Numerical simulation evaluating the total number of photons (laser + spontaneous emission) necessary for the vibrational cooling from a given $v_{X}$ toward the ground vibrational level ($v_{X}$ = 0).}
\label{photons}
\end{minipage}
\end{center}
\end{figure}

$ \nu_{laser}$ and $ \nu_{em}$ are, respectively, the laser frequency and the frequency of the spontaneously emitted photons. The broadband character of the laser enables the repetition of the pumping process from multiple vibrational $v_{X}$ levels. Depending on the wavelength, we observe a modification of the molecular resonance lines interpreted as a transfer of population between vibrational levels through optical pumping. By removing the laser frequencies corresponding to the excitation of a selected $v_{X}$ level, pumping molecules out of this level is made impossible such that it can be considered as a dark state. Progressively, the repetition of the absorption-spontaneous emission cycles lead to an accumulation of the molecules in the chosen $v_{X}$ level \cite{2009NJP}. In our first report on vibrational cooling by optical pumping cycles \cite{MatthieuViteau07112008} were sufficient to accumulate molecules into $v_{X}$ = 0. This result is confirmed by the present simulation. Taking intensity and spectral shape values close to the experimental ones, it evaluates the total number of photons (laser plus spontaneous emission), necessary for vibrational cooling from a given $v_{X}$ toward $v_{X}$ = 0. As shown in Figure \ref{photons}, roughly 10 photons, corresponding to the 5 laser cycles of our first experiment, are required to achieve vibrational cooling from level with ($v <$ 10).

\subsubsection{Laser shaping}
\begin{figure}
\begin{center}
\begin{minipage}{100mm}

\resizebox{8cm}{!}{{\includegraphics{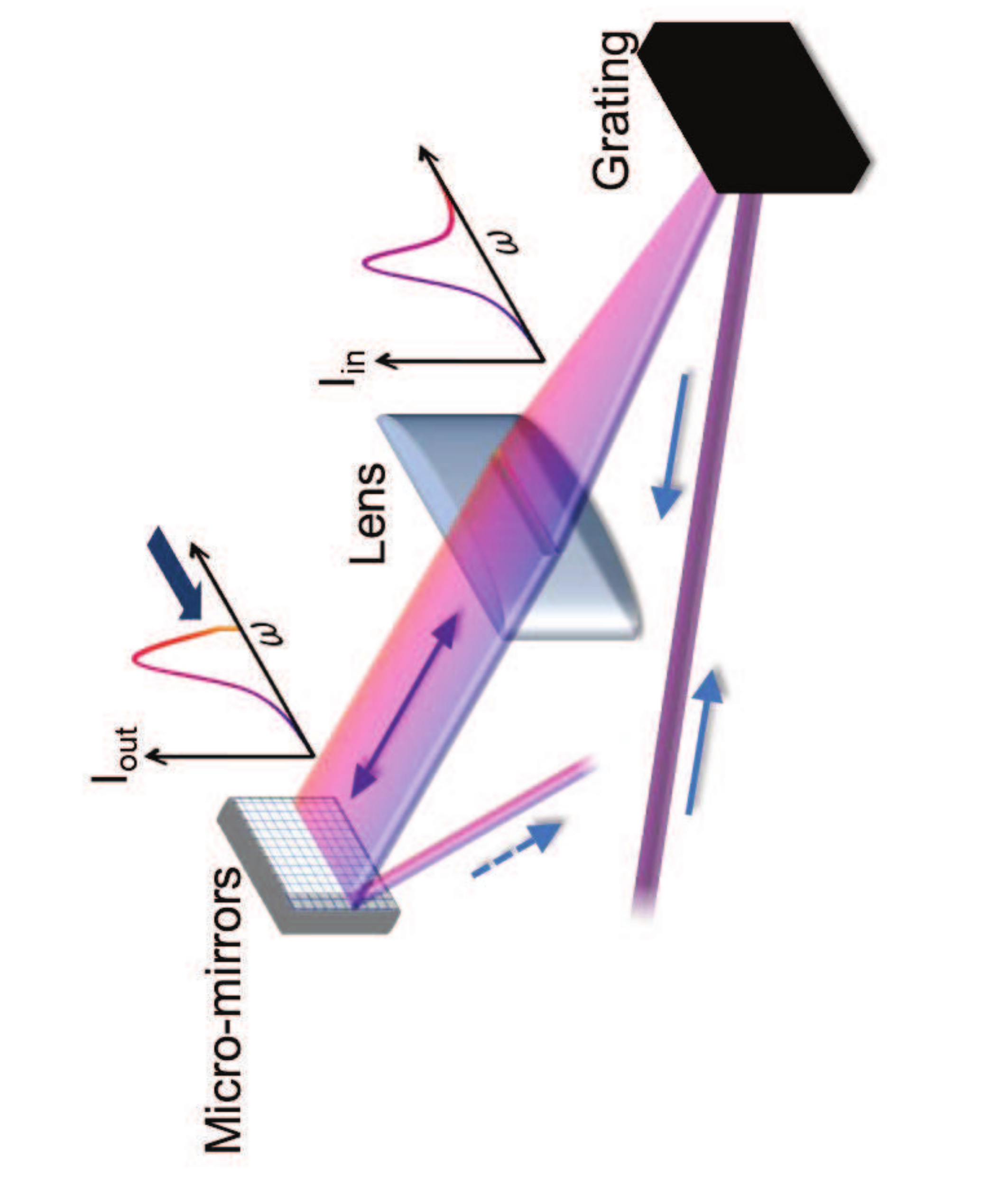}}}
\caption{Schematic representation of the 4-$f$ line used for the shaping of the broadband laser (from right to left : diffraction grating, cylindrical lens $f=100$mm, micro-mirrors) spatially spread the laser spectrum. ${\rm I}_{in}$ and ${\rm I}_{out}$ respectively sketch the incident laser beam and the reflected shaped laser beam.}
\label{4-fline}
\end{minipage}
\end{center}
\end{figure}

\begin{figure}
\begin{center}
\begin{minipage}{100mm}

\resizebox*{10cm}{!}{\includegraphics{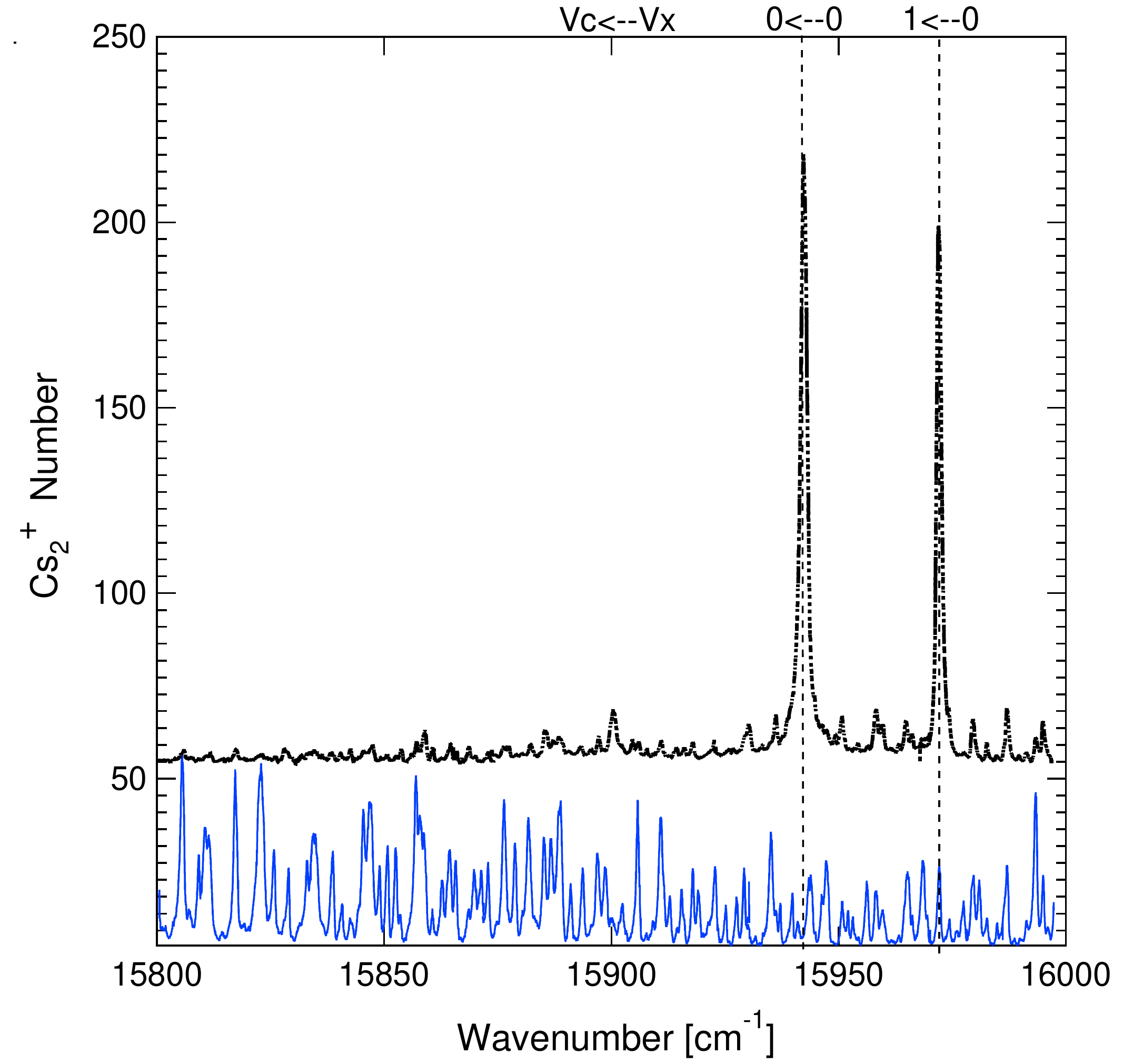}}
\caption{Photo-ionization spectra, without (blue spectrum shown in Fig.3) and with (black spectrum) vibrational laser-cooling. The vertical scale is the same for the two spectra and a constant of 50 ions offset is added to the black curve for clarity. Typically 80 \% of the total molecular population mainly distributed in the first vibrational levels is pumped to the ground vibrational level ($v$ = 0).}
\label{PIwithRV}
\end{minipage}
\end{center}
\end{figure}

The suppression of the optical frequencies to make dark a chosen state is obtained by spectral shaping the broadband laser. This shaping is realized by a simple 4-$f$ imaging system which is schematically represented in Figure \ref{4-fline}. The broadband laser is sent to a diffraction grating ($1800$ lines/mm) that spatially splits the different frequency components. The resulting beam is then collimated by a $100$ mm cylindrical lens and selectively reflected back by a mirror placed at focal length $f$ from the grating, i.e. in the Fourier plane of the optical system.  The laser spectrum being spatially spread over the whole mirror, a spatial mask or any means preventing the reflection locally leads to remove frequency components.
Ultimately, the frequency selection is obtained by using an array of rotatable micro-mirrors (DLP Texas Instruments Discovery .55 XGA 2xLVDS 120Type X Customer) that can be orientated according to two different angles. Thus, the frequency components necessary for the optical pumping are reflected back whereas those corresponding to the excitation from the dark state are rejected by a proper mirror tilt. It is noteworthy that if $v_X=0$ is the dark state, the shaping can be obtained by a simple cut off above $13030 \rm cm^{-1}$ in the laser spectrum (low-pass optical filter). As shown in Figure \ref{PIwithRV}, the application of the shaped laser strongly modifies the photo-ionization spectrum. Intense resonance lines reveal a net increase of population in $v_{X} = 0$ compared to the original PA spectrum.

The ranging applicability of this method exceeds the latter case and was demonstrated for many vibrational levels ($v_{X}$ = $1$, $2$ and $7$) by more complex laser shaping \cite{2009NJP}. The obvious limitation lies upon the available laser bandwidth and upon the initial molecular distribution: molecules reaching or initially lying in vibrational levels out of the range of the optical frequencies, escape from the pumping process.  

Several approaches can be implemented to optimize the spectral shaping. At some stage, they all require a simple criterion to keep or reject a given transition frequency. For example, a criterion that we often used was to choose the highest q$_{v', v''}$ factors between the target vibrational level and the excited states, and reject all the others. In general, it is thus useful to have access to the q$_{v', v''}$ factors to optimize the pumping process. In order to model the process, the optical cycle is divided in two steps: the excitation (absorption) from X to B, and the spontaneous emission B to X. In a perturbative approach (low laser intensity), the distribution of molecules in the various vibrational levels is described in absorption by  

\begin{eqnarray*}
\begin{aligned}
{\rm P}_{v_{\rm X}}^{(n+1)} &= {\rm P}_{v_{\rm X}}^{(n)} - \sum \limits_{\substack{v_{\rm B}}} A_{v_{\rm B}v_{\rm X}} {\rm P}_{v_{\rm X}}^{\it{(n)}},\\
{\rm P}_{v_{\rm B}}^{(n+1)} &= \sum\limits_{\substack{v_{X}}} {\rm A}_{v_{\rm B}v_{\rm X}} {\rm P}_{v_{\rm X}}^{(n)},
\end{aligned}
\end{eqnarray*}

and in emission by 

\begin{eqnarray*}
{\rm P}_{v_{\rm X}}^{(n+2)} = {\rm P}_{v_{\rm X}}^{(n+1)} + \sum \limits_{\substack{v_{\rm B}}} E_{v_{\rm X}v_{\rm B}} {\rm P}_{v_{\rm B}}^{(n+1)}.
\end{eqnarray*}

P$_{v_{\rm X}}^{(n)}$ and P$_{v_{\rm B}}^{(n)}$ respectively denote the population of the vibrational levels $v_{\rm X}$ and $v_{\rm B}$ of the ground and the excited electronic states at the stage number $n$ of the process. A$_{v_{\rm X}v_{\rm B}}$ denotes the ratio at which molecular populations are transferred from one vibrational level $v_{\rm X}$ to any $v_{\rm B}$ accessible by our laser bandwidth. Similarly, E$_{v_{\rm B}v_{\rm X}}$ denotes the molecular population decay by spontaneous emission from one vibrational level $v_{\rm B}$ to $v_{\rm X}$. These rates are proportional to :

\begin{eqnarray*} 
\begin{aligned}
& {\rm A}_{v_{\rm X}v_{\rm B}} \propto {\rm q}_{v_{\rm X}v_{\rm B}}({\rm D}_{v_{\rm X}v_{\rm B}})^{2} {\rm I}_{{\rm laser}, v_{\rm X}v_{\rm B}}, \\
& {\rm E}_{v_{\rm X},v_{\rm B}} \propto {\rm q}_{v_{\rm X}v_{\rm B}}({\rm D}_{v_{\rm X}v_{\rm B}})^{2} \omega^{3}_{v_{\rm X}v_{\rm B}}.\\
\end{aligned}
\end{eqnarray*}

${\rm q}_{v_{\rm X}v_{\rm B}}$, $({\rm D}_{v_{\rm X}v_{\rm B}})^{2}$ and $\omega_{v_{\rm X}v_{\rm B}}$ correspond to the Franck-Condon coefficient, the dipole moment and the transition frequency of the vibrational transition, respectively. As a consequence, after N steps of absorption emission cycles,

\begin{eqnarray*}
{\rm P}^{(2N)}= {\rm M}^{N}{\rm P}^{(0)},
\end{eqnarray*}
with  M$_{ij} = (1 + {\rm EA})_{ij} - \sum\limits_{\substack{k}} {\rm A}_{ki}\delta_{ij} = \delta_{ij} + \sum\limits_{\substack{k}} ({\rm E}_{ik}{\rm A}_{kj}- {\rm A}_{ki}\delta_{ij})$. \\
For example, a very simple optimization procedure to pump toward a single level $v_{0}$ is to set the laser intensity (${\rm I}_{{\rm laser},v_{\rm X}v_{\rm B}}$) either ON or OFF. We can then write ${\rm A}_{ij} = {\rm A}_{ij}\Delta_{ij}$ where $\Delta_{ij} = 0$ for laser OFF and $\Delta_{ij} = 1$ for laser ON. The criterion is that P$^{(2)}$ has to get closer to P$_{0}$, a target distribution defined by $({\rm P}_{0})_{i}= \delta_{i v_{0}}$. It means that we have to minimize the L$_{1}$ norm 

\begin{eqnarray*}
\begin{aligned}
\parallel {\rm P}^{(2)}- {\rm P}_{0}\parallel_{1} &= 1- {\rm P}_{v_{0}}^{(2)} + \sum\limits_{\substack{v_{\rm X}\neq v_{\rm 0}}}{\rm P}^{(2)}_{v_{\rm X}}, \\
&= 1-{\rm P}^{(2)}_{v_{0}} + \sum\limits_{\substack{v_{\rm X}\neq v_{0}}}P_{v_{\rm X}}^{(0)} + \sum\limits_{\substack{v_{\rm B}v_{\rm X'}}} \Delta_{v_{\rm B}v_{\rm X'}} {\rm A}_{v_{\rm B}v_{\rm X'}}{\rm P}_{v_{\rm X'}}^{(0)} \sum\limits_{\substack{v_{\rm X}\neq v_{0}}}({\rm E}_{v_{\rm X}v_{\rm B}}-\delta_{v_{\rm X}v_{\rm X'}}).
\end{aligned}
\end{eqnarray*}

Noting that if $v_{\rm X'}=v_{0}$, $\sum\limits_{\substack{v_{\rm X}\neq v_{0}}}({\rm E}_{v_{\rm X}v_{\rm B}}-\delta_{v_{\rm X}v_{\rm X'}})>0$, we just have to choose $\Delta_{v_{\rm B}v_{\rm 0}} = 0$. Therefore, $v_{0}$ is a dark state and cannot interact with the laser. But if $v_{\rm X'}\neq v_{0}, \sum\limits_{\substack{v_{\rm X}\neq v_{0}}}({\rm E}_{v_{\rm X}v_{\rm B}}-\delta_{v_{\rm X}v_{\rm X'}})< 0$, and then the minimization of $\parallel {\rm P}^{(2)}- {\rm P}_{0} \parallel$ implies that $\Delta_{v_{\rm B}v_{\rm X'}} = 1$, i.e. the laser is turned ON to excite all the population lying in vibrational levels $v_{\rm X'}\neq v_{0}$. 

\section{Rotational Spectroscopy}
\subsection{Detection of rotation}
In principle, the rotational distribution of cold molecules could be observed also without vibrational cooling but, after photoassociation, only a small number of molecules is produced in any given vibrational level $v_{\rm X}$. It would be thus very difficult to resolve spectroscopically the rotational structure. That's why we apply a vibrational cooling phase before attempting to detect the rotation.   
When molecules are transferred to the ground vibrational level ($v=0$), they are still spread over several (typically 4-6) rotational levels. 
The rotational splitting of Cs$_2$ levels is too narrow to be resolved with a pulsed dye laser because the linewidth is larger compared to the separation between two rotational levels ($\sim 600$ MHz). Other methods must be adopted to perform a rotational spectroscopy of our molecular sample. They are based on using a continuous narrow-band laser that can selectively excite a rotational level $J_{\rm X}$.
The simplest detection method would be a direct ionization of the $v_{B}$ population after a very fast narrow-band laser pulse according to : 

\begin{eqnarray*}
\begin{aligned}
& {\rm Cs}_2(v_{\rm X} = 0, J_{\rm X}) + h \nu_{\rm laser\ diode} \rightarrow {\rm Cs}_2^*(v_{\rm B}, J_{\rm B})\ {\rm (selective\ excitation)},\\
& {\rm Cs}_2^*(v_{\rm B}, J_{\rm B}) + h \nu_{\rm pulsed \ laser} \rightarrow {\rm Cs}_{2}^{+} {\rm (ionization)}.
\end{aligned}
\end{eqnarray*}

However, it is so tricky to generate such a short pulse of the required narrow-band laser, that we have used other detection methods. They are less obvious but more efficient. 

\subsubsection{Depletion spectroscopy} 
\begin{figure}
\begin{center}
\begin{minipage}{10cm}
\centering{\subfigure{
(a)\resizebox*{9cm}{!}{\includegraphics{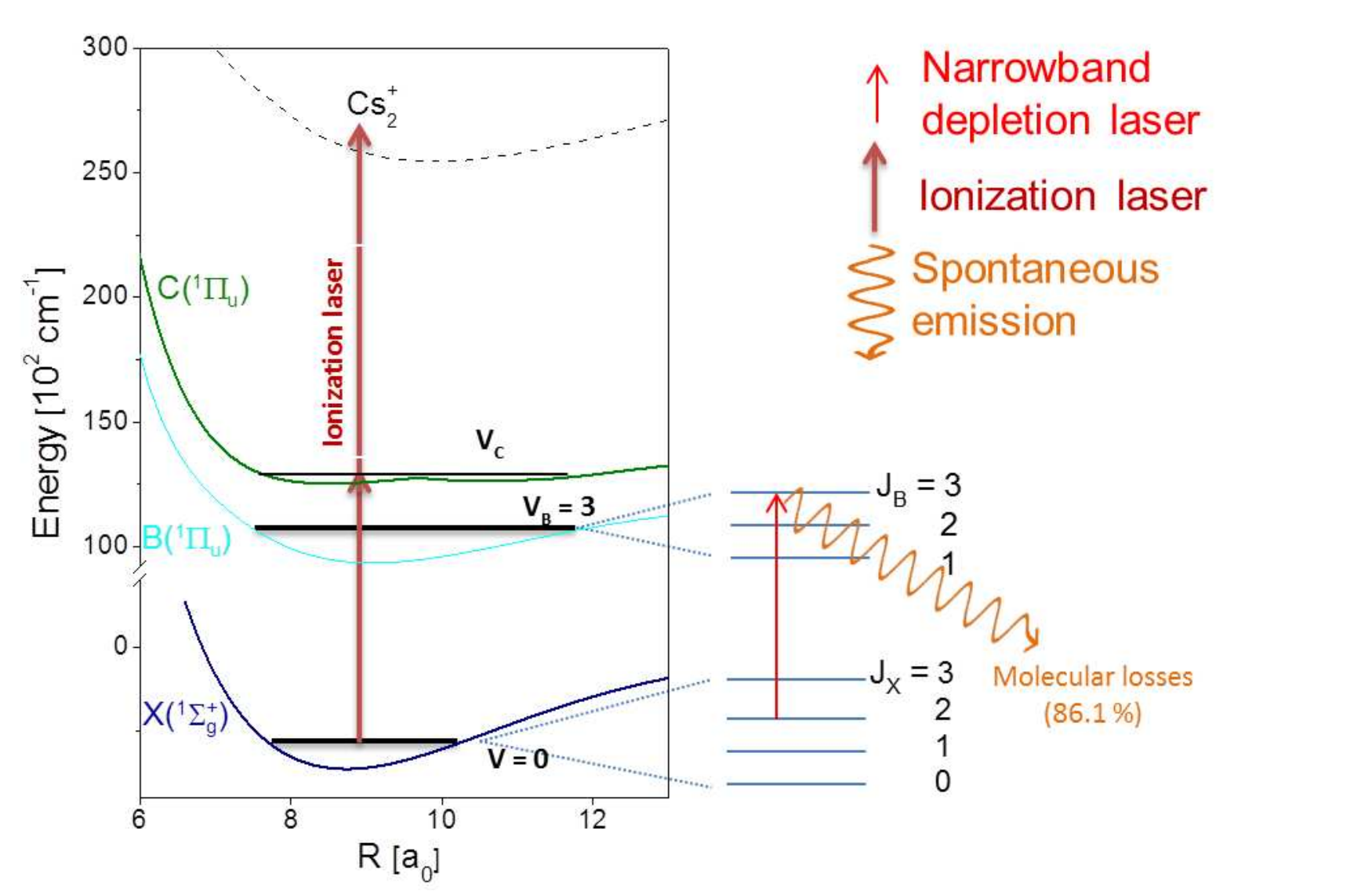}}}
\subfigure{
(b)\resizebox*{9cm}{!}{\includegraphics{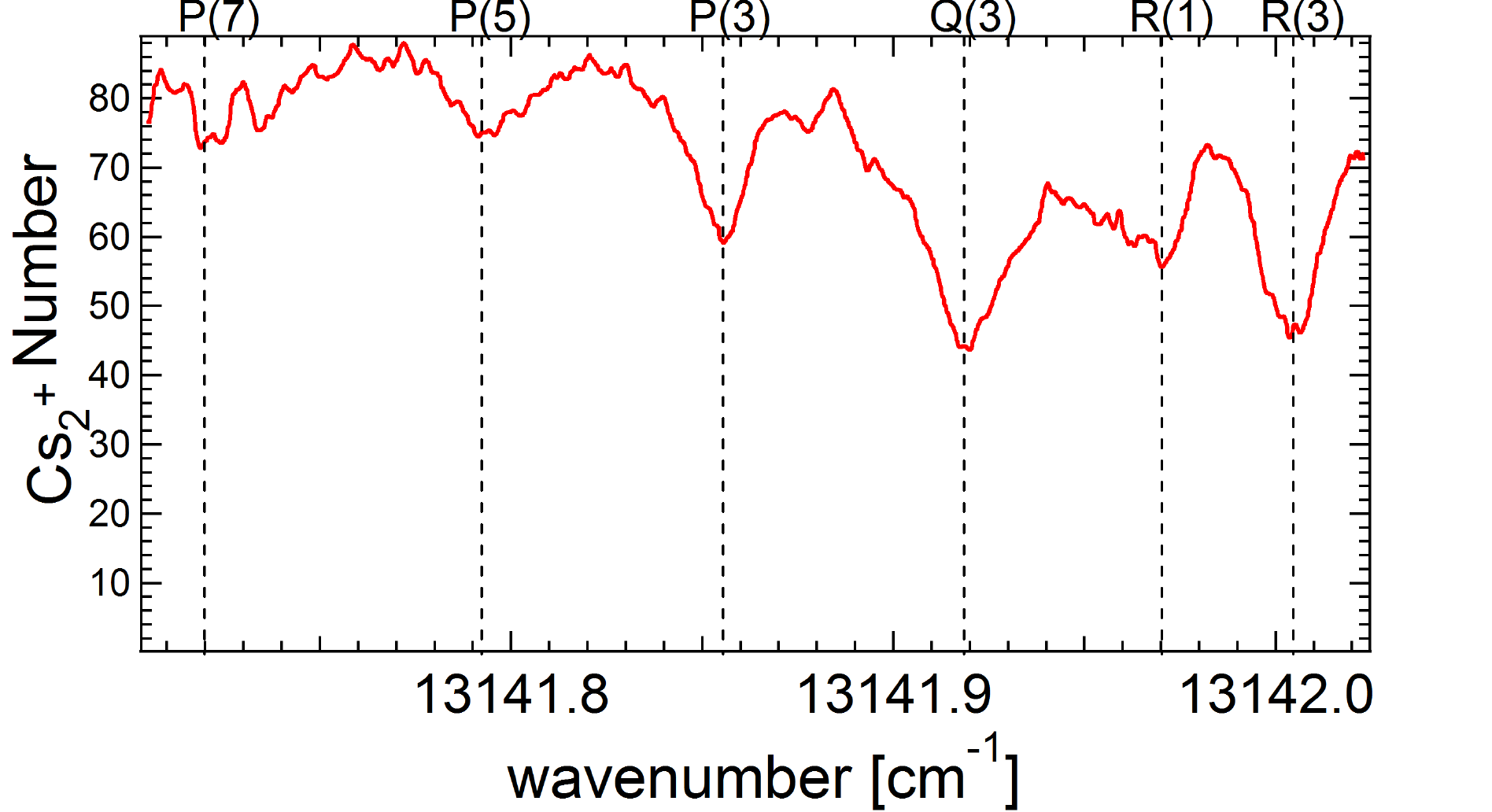}}}}
\caption{(a) Schematic representation of the transitions involved to resolve the rotational spectrum by the depletion spectroscopy method. (b) In red, depletion spectrum of ground state $v_{X} = 0$ level. The depletion laser is scanned over $(v_{\rm B} = 3,J_{\rm B}) \leftarrow (v_{\rm X} = 0, J_{\rm X})$ rotational transitions.}
\label{sample-figure}
\end{minipage}
\end{center}
\end{figure}

The idea of the depletion spectroscopy \cite{Wang07} is to use a continuous-wave (cw) diode laser to deplete molecules from the ground vibrational level, while the loss is monitored by a REMPI technique. In our experiment, we used a cw DFB laser scanned over the $B^{1}\Pi_{u}(v_{\rm B}=3, J_{\rm B}) \leftarrow {\rm X}^1\Sigma^{+}_{g}(v_{\rm X}=0,J_{\rm X})$ transitions as schematically shown in Figure \ref{sample-figure}a). The Franck-Condon factors of this transition are not particularly strong but this choice was imposed by the wavelength of the DFB laser available in our laboratory. The population in $v=0$ evolves according to the following equations

\begin{eqnarray*}
\begin{aligned}
& {\rm Cs}_2(v_{\rm X}=0, J_{\rm X}) +h\nu_{\rm DFB} \rightarrow {\rm Cs}_2^*(v_{\rm B}=3, J_{\rm B}=J_{\rm X}\pm 0,1),\\
& {\rm Cs}_2(v_{\rm X}=0)+  2 h\nu_{\rm pulsed \ laser} \rightarrow {\rm Cs}_2^*(v_{C}) + h \nu_{\rm pulsed \ laser} \rightarrow {\rm Cs}_{2}^{+}\ {\rm (REMPI\ detection)}.\\
\end{aligned}
\end{eqnarray*}
A typical example of a depletion spectrum, obtained after 50 $\mu$s irradiation with the DFB laser is shown in Figure \ref{sample-figure}b). 
 
\subsubsection{Spontaneous-decay-induced double resonance (SpIDR)}
\begin{figure}
\begin{center}
\begin{minipage}{100mm}
\centering{
\resizebox*{10cm}{!}{\includegraphics{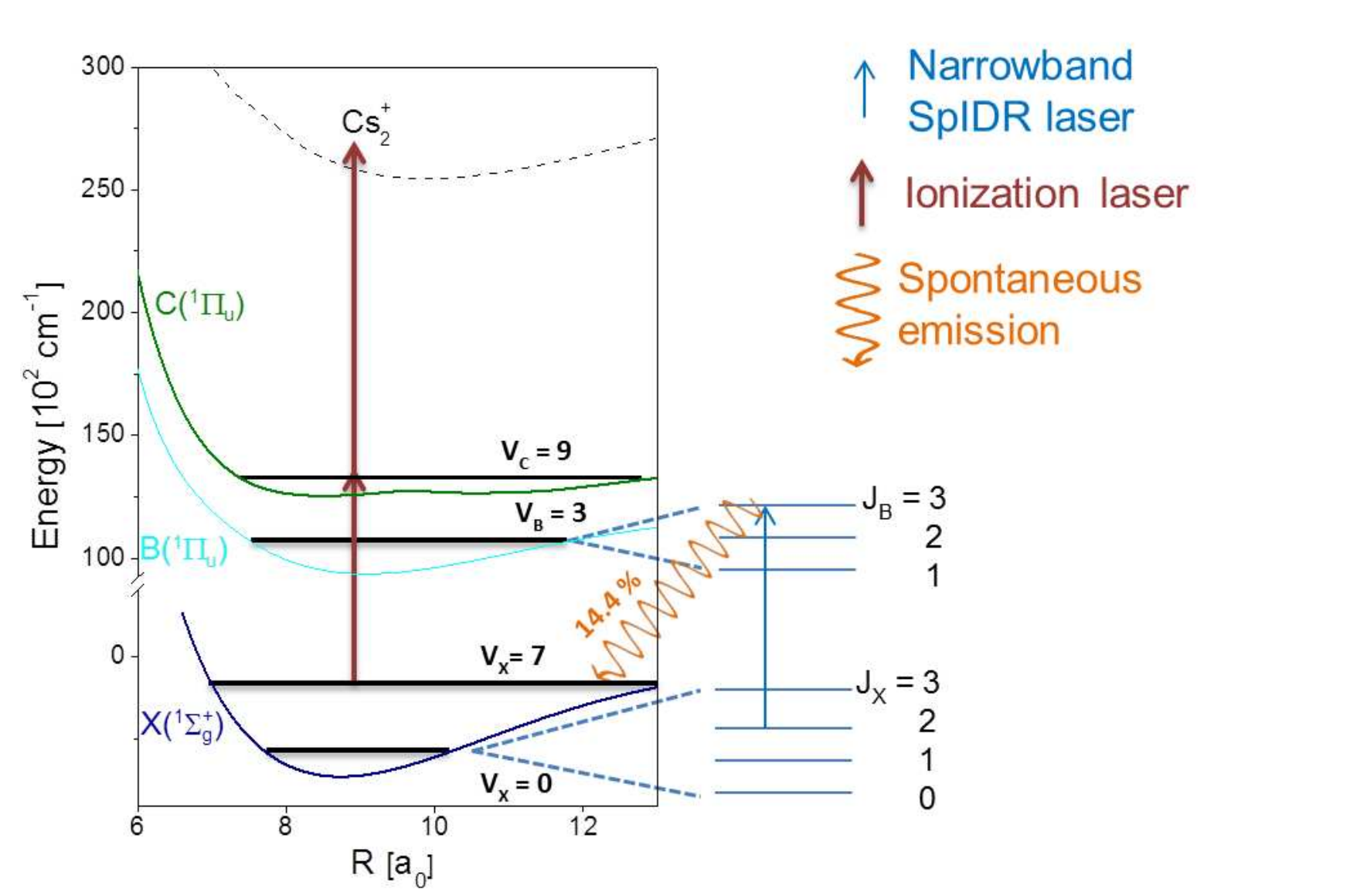}}}
\caption{(color online) Transitions used to detect the rotational spectrum after the optical pumping toward the ground vibrational level. The rotational spectroscopy is performed by a narrowband SpIDR laser which is scanned over ($v_{\rm B}$ = 3, J$_{\rm B}$) $\leftarrow$ ($v_{\rm X}$ = 0, J$_{\rm X}$) rotational transitions. The brown arrow represents the pulsed laser that ionizes molecules in the $v_{\rm X}$ = 7 level, through the intermediate C$^{1}\Pi_{u}$ state ($v_{\rm C}$ = 9).}
\label{spidr}
\end{minipage}
\end{center}
\end{figure}

The second detection method we used relies on measuring the spontaneous-decay-induced signal of a vibrational level of the ground electronic state having a good Franck-Condon $q$ factor with a vibrational level of the excited electronic state. In our experiment, $v_{\rm X}=7$ has been chosen because its $q$ factor with $v_{\rm B}=3$ is the largest available one ($\sim 0.15$) as shown in Table 1, and because $v_{\rm X}=7$ is efficiently emptied by the vibrational optical pumping. 
The $v_{\rm X}=7$ population is probed by the pulsed dye laser through the $v_{\rm C}=9 \leftarrow  v_{\rm X}=7$ (vibrational transition). The various processes involved are

\begin{eqnarray*}
\begin{aligned}
& {\rm Cs}_2(v_{\rm X}=0, J_{\rm X})+h \nu_{\rm DFB} \rightarrow {\rm Cs}_2^*(v_{\rm B}=3, J_{\rm B})\ \rm{(excitation)},\\
& {\rm Cs}_2^*(v_{\rm B}=3, J_{\rm B}) \rightarrow {\rm Cs}_2(v_{\rm X}=7, J_{\rm X}) + h \nu_{em}\ (\rm{spontaneous\ emission}),\\
& {\rm Cs}_2(v_{\rm X}=7) + 2 h\nu_{\rm pulsed \ laser} \rightarrow {\rm Cs}_2^*(v_{\rm C})+ h \nu_{\rm pulsed \ laser} \rightarrow {\rm Cs}_{2}^{+}\ {\rm (REMPI\ detection)}.\\
\end{aligned}
\end{eqnarray*}
The scheme of the laser transitions used to detect the rotation with SpIDR method are shown in Figure \ref{spidr}. The main difference with the preceding depletion method is in the pulsed laser wavelength.

As in reference \cite{Aikawa10}, we found that this method requires smaller power and shorter pulse duration than depletion spectroscopy. This may look surprising because those techniques only differ by the post-ionization step. The reason is that the spontaneous-decay-induced signal is almost background free whereas the depletion signal starts from the fluctuating signal of the $ v_{X}=0$ population which strongly alters the signal-to-noise ratio. A first example of spectrum acquired by the spontaneous-decay-induced signal method is shown in Figure \ref{sample-figure1}.

\subsection{Spectrum analysis} 
\begin{figure}
\begin{center}
\begin{minipage}{10cm}
\subfigure{
\resizebox*{10cm}{!}{\includegraphics{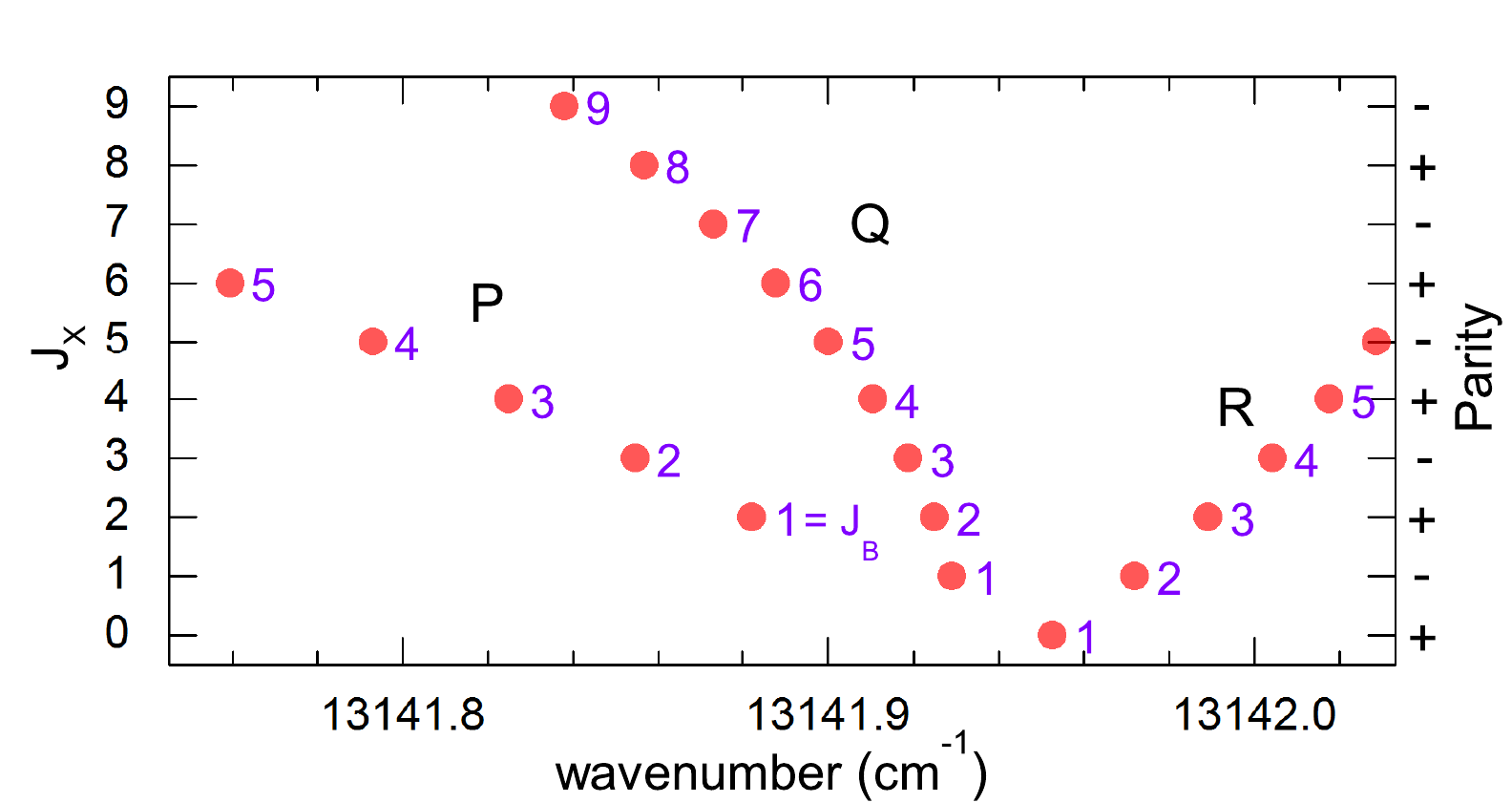}}}
\subfigure{
\resizebox*{10cm}{!}{\includegraphics{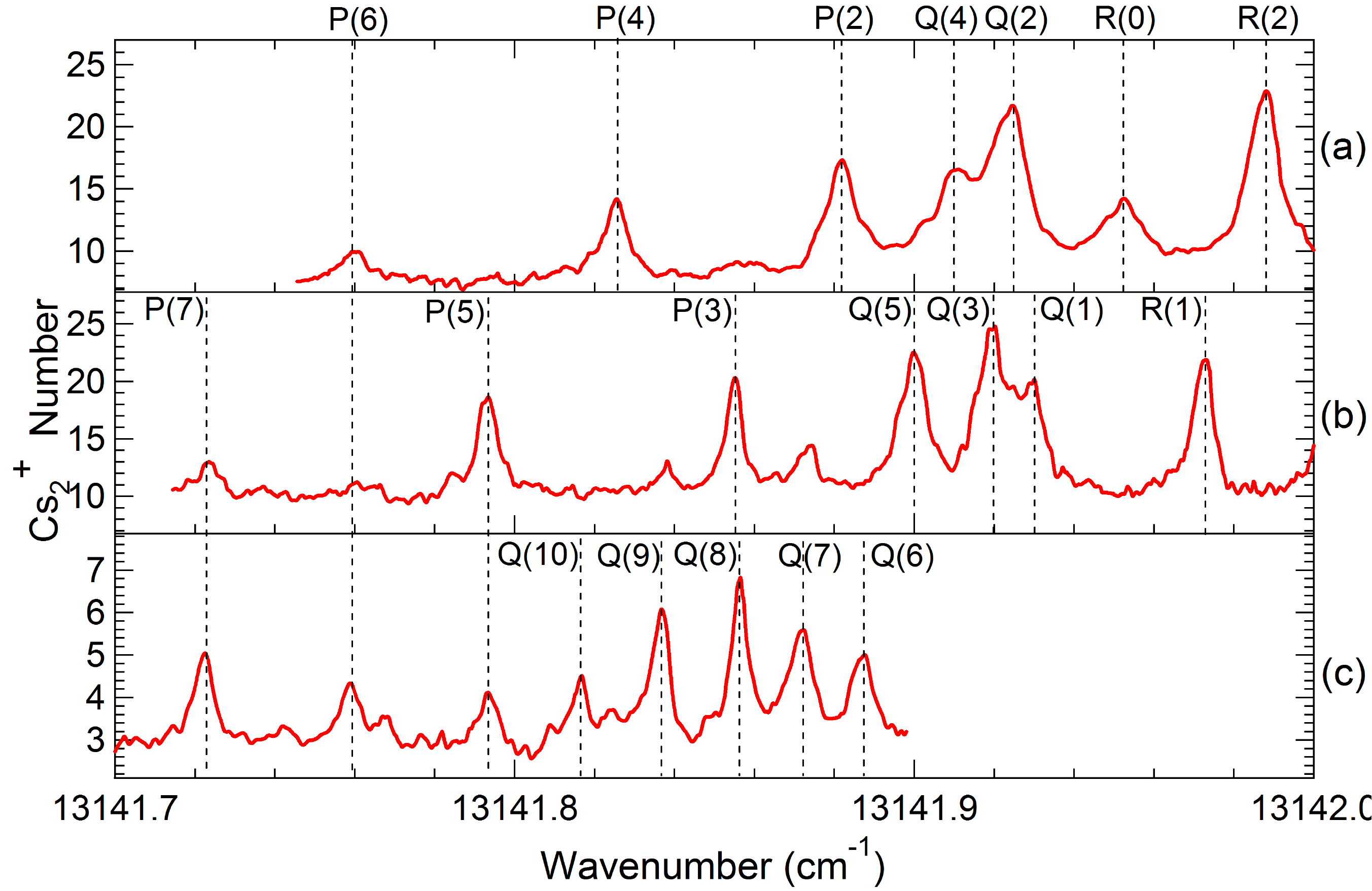}}}
\caption{Upper panel: Fortrat diagram showing the transition wavenumbers from the ($v=0,\ 0\leqslant J \leqslant9$) rotational levels of the X state. The P, Q, and R branches, correspond to the $J_{\rm B} - J_{\rm X} = -1, 0\ {\rm and}\ 1$ allowed transitions. The parity of the $J$ state is also indicated on the right-hand side. The lower panel shows the rotational distribution in the ground vibrational level (X$^{1}\Sigma_{g}^{+}$, $v_{\rm X} = 0$) after vibrational cooling has been applied} with three different PA schemes: through a) $0_g^-$ $(v=26, J=1)$, b) ${\rm }0_g^-(v=26, J=2)$ and c) $1_g$ $(v, {\rm J=8})$. The two first spectra reveal lines of a unique $J_{\rm X}$ parity.
\label{sample-figure1}
\end{minipage}
\end{center}
\end{figure}

As described above, photo-association through the ${\rm 0}_{g}^{-}(6s + 6p_{1/2})$ state, leads to the formation of cold molecules in the X ground electronic state by two-photon decay through ${\rm 0}_{u}^{+}$ state. After the PA process, we apply the vibrational optical pumping to cool the vibration. When molecules are optically pumped in the ground vibrational level ($v=0$) we detect the rotation distribution of molecules in this level. We then obtain the rotational spectrum shown in the lower panel of Figure \ref{sample-figure1}. We note that only even or odd rotational states are populated. It is a strong proof that the cold molecule formation process is indeed two-photons decay. It can be easily explained by the fact that the rotational states have a (+/-) parity given by the sign of $(-1)^{J+1}$ in ${\rm 0}_g^-$ excited state and $(-1)^J$ in the ground electronic state (X). The parity conservation is also a piece of evidence that the vibrational optical pumping process using the B state does not change the parity \cite{HERZBERG}. This also means that the two photon decay through $0_{u}^{+}$ state is not selective in parity, which is expected. We can also obtain spectra with both even and odd rotational states as shown in Figure \ref{sample-figure1}c) if $1_g$($6s+6p_{3/2}$) is used for the PA process. This is possible because 1$_{g}$ state has no definite parity. We also note that, in this case, the rotational distribution is centered around higher $J$ values, as the chosen PA level is $J$ = 8. 

\begin{figure}
\begin{center}
\begin{minipage}{100mm}
\centering{
\resizebox*{7cm}{!}{\includegraphics{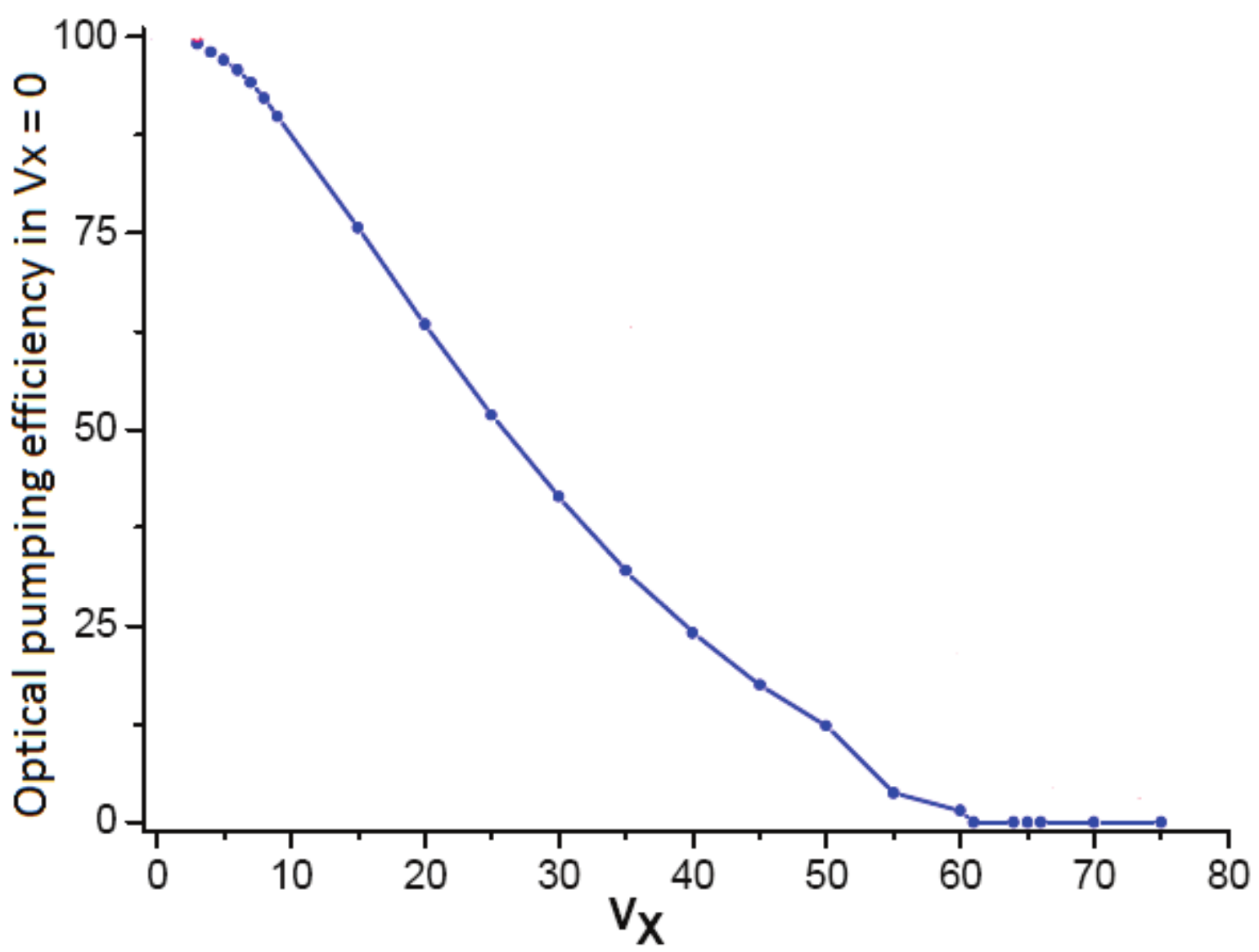}}}
\caption{Numerical simulation evaluating the transfer efficiency of molecules initially in a given $v_{\rm X}$ toward the ground vibrational $v_{\rm X} =0$.}
\label{evolutionpopulation}
\end{minipage}
\end{center}
\end{figure}

\section{Rotational cooling}
\begin{figure}
\begin{center}
\begin{minipage}{100mm}
\centering{
\resizebox*{8cm}{!}{\includegraphics{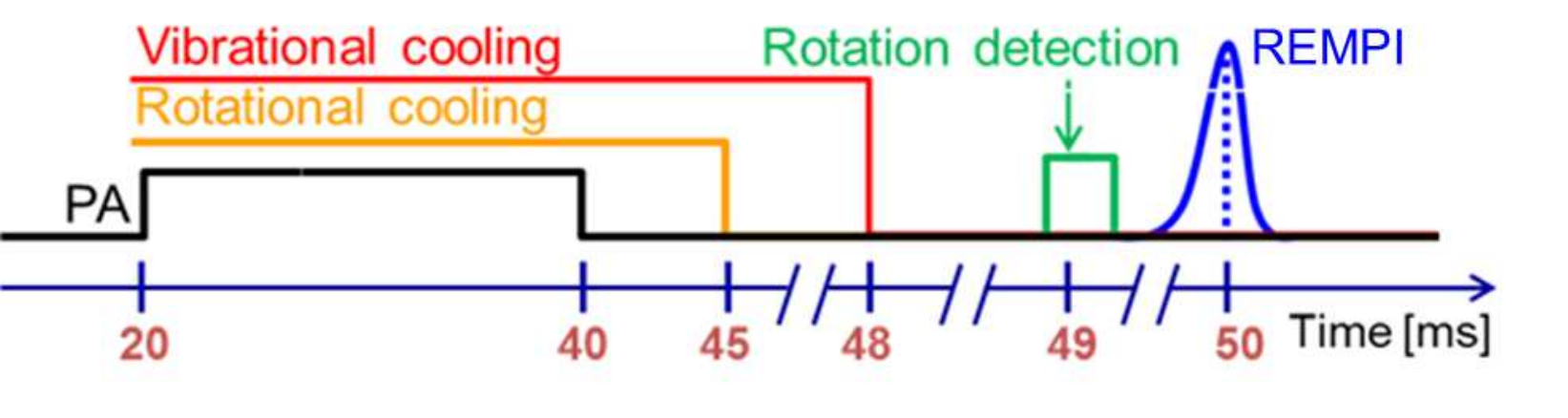}}}
\caption{(color online) Temporal sequence used to cool the rotation and vibration : PA is applied for $\rm 20ms$, the rotational and vibrational cooling lasers are applied simultaneously but they are turned off respectively $5\ \rm ms$ and $8 \rm ms$ after the PA laser. $1$ ms later a fourth laser is used, during typically $50\ \umu \rm{s}$, for detection of the rotational population. Finally 1 ms later a pulsed Nd:YAG laser, is used to excite ${\rm X}^{1}\Sigma^{+}_{g}$ vibrational level, via the ${\rm C}^{1}\Pi_{u}$ or ${\rm D}^{1}\Sigma_{u}^{+}$ states in order to produce ions that can be detected by the micro-channel plates (MCPs).}
\label{temporelsequence}
\end{minipage}
\end{center}
\end{figure}

\begin{figure}
\begin{center}
\begin{minipage}{100mm}
\centering{
\resizebox*{8cm}{!}{\includegraphics{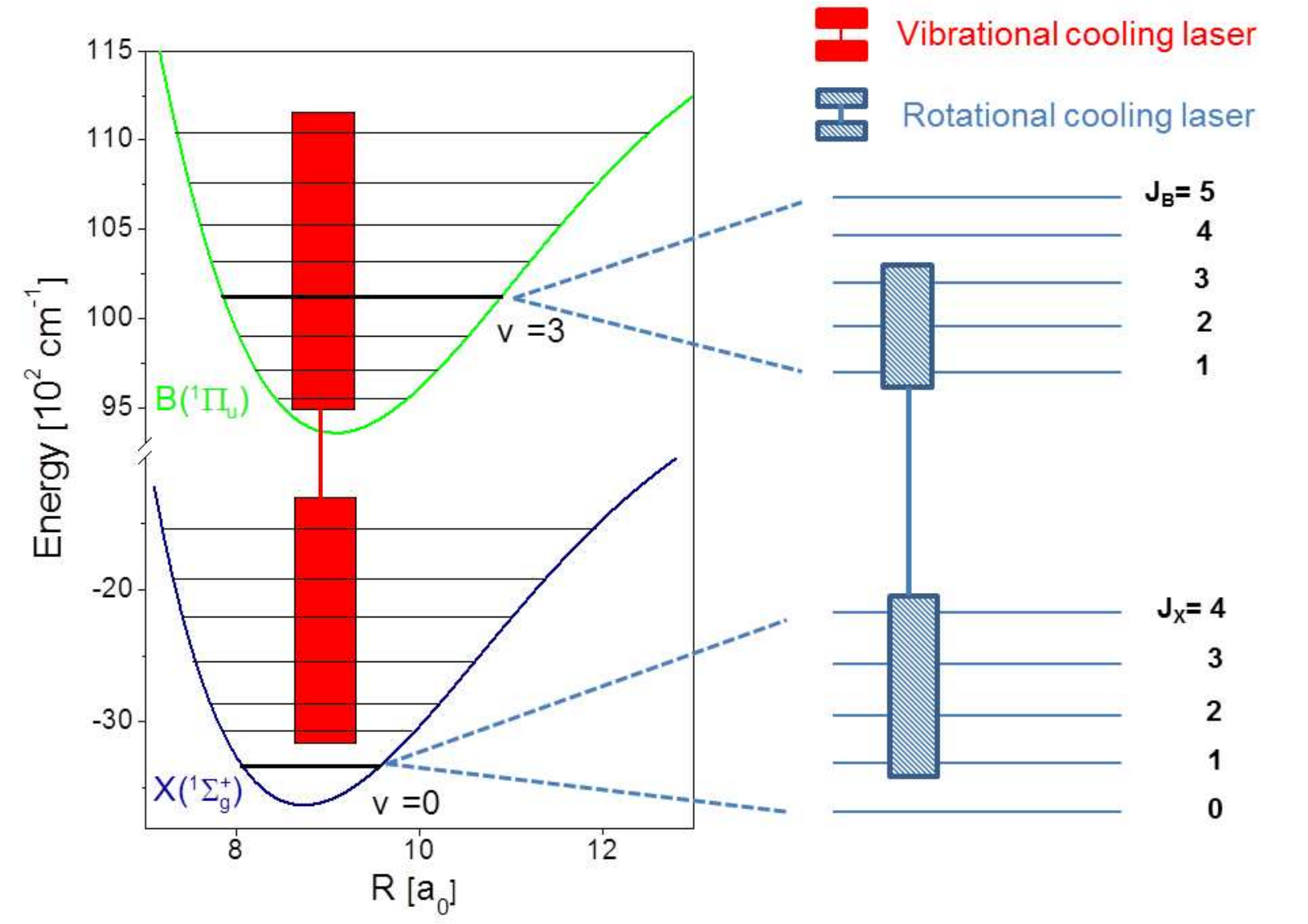}}}
\caption{Transitions used to cool the vibration and rotation of $\rm Cs_{2}$ molecules. The red filled rectangle and (blue) hatched rectangle correspond respectively to the vibrational levels affected by the laser cooling of the vibration and the rotational states affected by the temporal scanning of the narrowband diode laser. The upper and lower numbers on the right hand side are $\rm J_{B}$ of $v_{B} = 3$ and J of the ground vibrational level.}
\label{refdet}
\end{minipage}
\end{center}
\end{figure}

Photoassociation on the $0_{g}^{-}(6s + 6p_{1/2})$ state leads to significant population in the ground electronic state but spread out in more than $60$ vibrational levels. In order to observe the rotational population we need to increase the population in a given vibrational state. In Ref. \cite{2011PCCP...1318910L} we have shown that the vibrational pumping toward $v = 0$ is very efficient only for low vibrational levels ($v_{\rm X} < 10$). This result is confirmed by computing, via a numerical simulation, the pumping efficiency from a given $v_{\rm X}$ toward $v_{\rm X} =0$ as shown in Figure \ref{evolutionpopulation}.
This threshold is only due to the limited laser bandwidth but implies that only 25\% of the initial distribution is pumped into the $v_{\rm X} =0$ level. As already evoked, molecules finally lying in ($v_{\rm X} =0$) are spread over $\sim$  four $J_{\rm X}$ states (Figures \ref{sample-figure} and \ref{sample-figure2}). The manipulation of the rotational populations is based on the same principle as that used to achieve the vibrational cooling scheme : a broadband laser excites all the rotational states except the one of interest where the molecules accumulate after spontaneous emission. However, in the case of a heavy molecule like Cs$_{2}$, the rotational spectrum is too narrow to be resolved using a simple grating, and thus the resolution limitation must be improved. To do so, two distinct laser sources are actually simultaneously used. The first one, devoted to the vibrational cooling, is a broadband laser with its spectrum shaped by a grating and a selective element in the Fourier plane. The second one is a narrow-bandwidth DFB diode laser, whose spectrum is artificially broadened by modulating the diode current to obtain a $\sim$ 5 GHz uniform scan range. The amplitude, center, speed and repetition of the scan is controlled by a function generator driving the laser diode current. The effect of temporal scanning is identical to that an broadband source with the same width. The narrow-band cw diode laser, ($ \sim 2$ MHz, $ \sim 6$ mW power, 3.5 mm$^{2}$ beam size), is continuously scanned across the proper ${\rm B}^1\Pi_u ( v_{\rm B}=3, {J_{\rm B}}) \leftarrow {\rm X} ^1\Sigma_g^+(v_{\rm X}=0, J_{\rm X})$ transitions with a period of 100 $\mu$s. As the rotational cooling laser tends to redistribute the molecules in other vibrational levels due to unfavorable Franck-Condon factors \cite{Fioretti09}, the rotational cooling requires a simultaneous vibrational phase. By studying the $q$ factors, we know that the spontaneous decay from $v_{B}=3$ mostly populates  $v_{\rm X}< 10$, which provides a global quasi-closed pumping system. The relevant Franck-Condon factors between the $v_{B} = 3$ and $v_{X} (0-10)$ levels are shown in Table 1. A schematic presentation of the rotational cooling laser scheme is shown in Figure \ref{refdet} when the rotational cooling laser is tuned between the rotational levels J$_{X}$ of the ground vibrational state and the rotational levels of the excited vibrational level $v_{B}=3$. A typical temporal sequence allowing for PA and ro-vibrational cooling is reported in Figure \ref{temporelsequence}.

\begin{table}
\tbl{Franck-Condon factors (x 100) between B$^{1}\Pi_{u}$ ($v'$ = 3) and X$^1\Sigma_g^+$ ($v''$ = 0-10) \cite{Diemer89}}
{\begin{tabular}{@{}lccccccccccc}\toprule
 X$^1\Sigma_g^+$& $v$=0& $v$=1& $v$=2& $v$=3& $v$=4& $v$=5& $v$=6& $v$=7 &$v$=8 &$v$=9 &$v$=10\\
\colrule
 q$_{v', v''}$& 13.9& 17.7& 2.8& 7.7& 10.8& 0.4& 5.4& \textbf{14.4} &14 &8.2&3.4\\
\botrule
\label{FCtable} 
\end{tabular}}
\end{table}

\begin{figure}
\begin{center}
\begin{minipage}{100mm}
\centering{
\resizebox*{8cm}{!}{\includegraphics{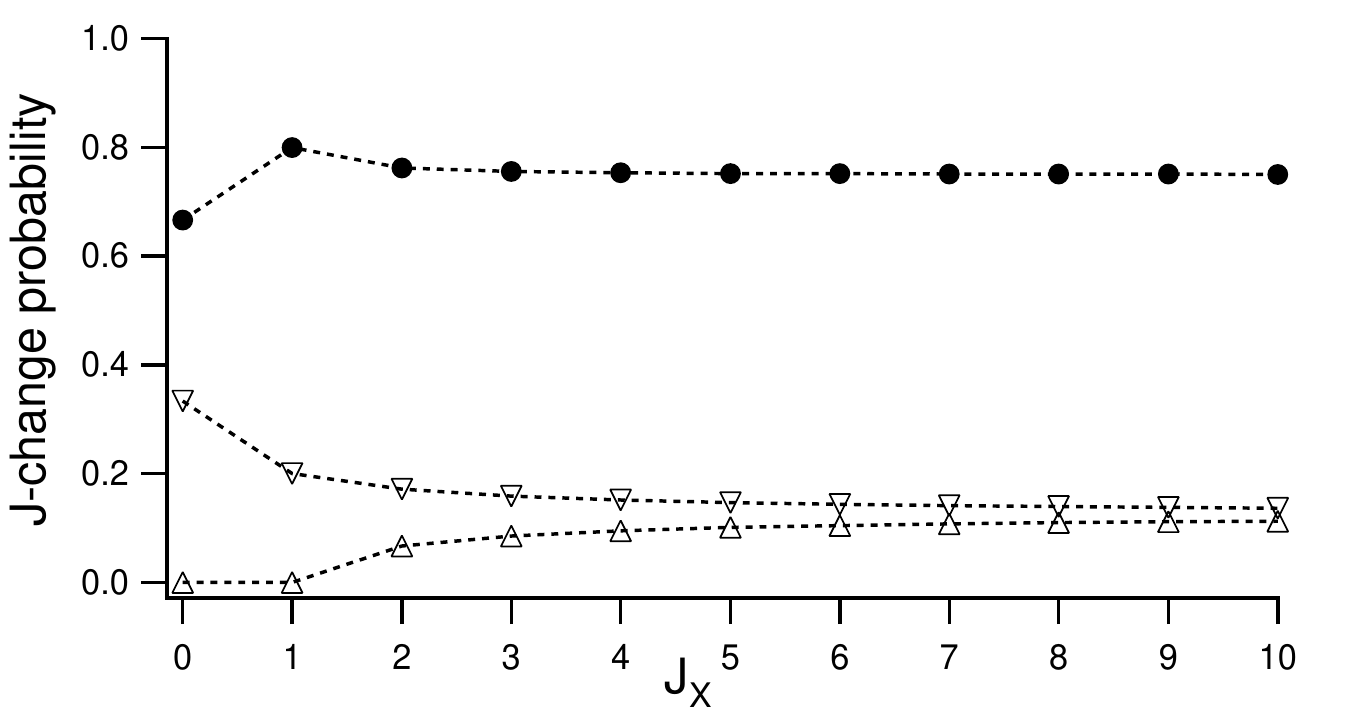}}}
\caption{Effect of vibrational cooling on the rotational distribution : simple Hönl-London formula give the probability to find $J_X-2$ (down triangles), $J_X$ (circles), $J_X+2$ (up triangles) from a given $J_X$ after a cycle of absorption/spontaneous emission produced by vibrational cooling.}
\label{Jchange_vibcooling}
\end{minipage}
\end{center}
\end{figure}

\begin{figure}
\begin{center}
\begin{minipage}{\textwidth}
\subfigure{
\resizebox*{7cm}{!}{\includegraphics{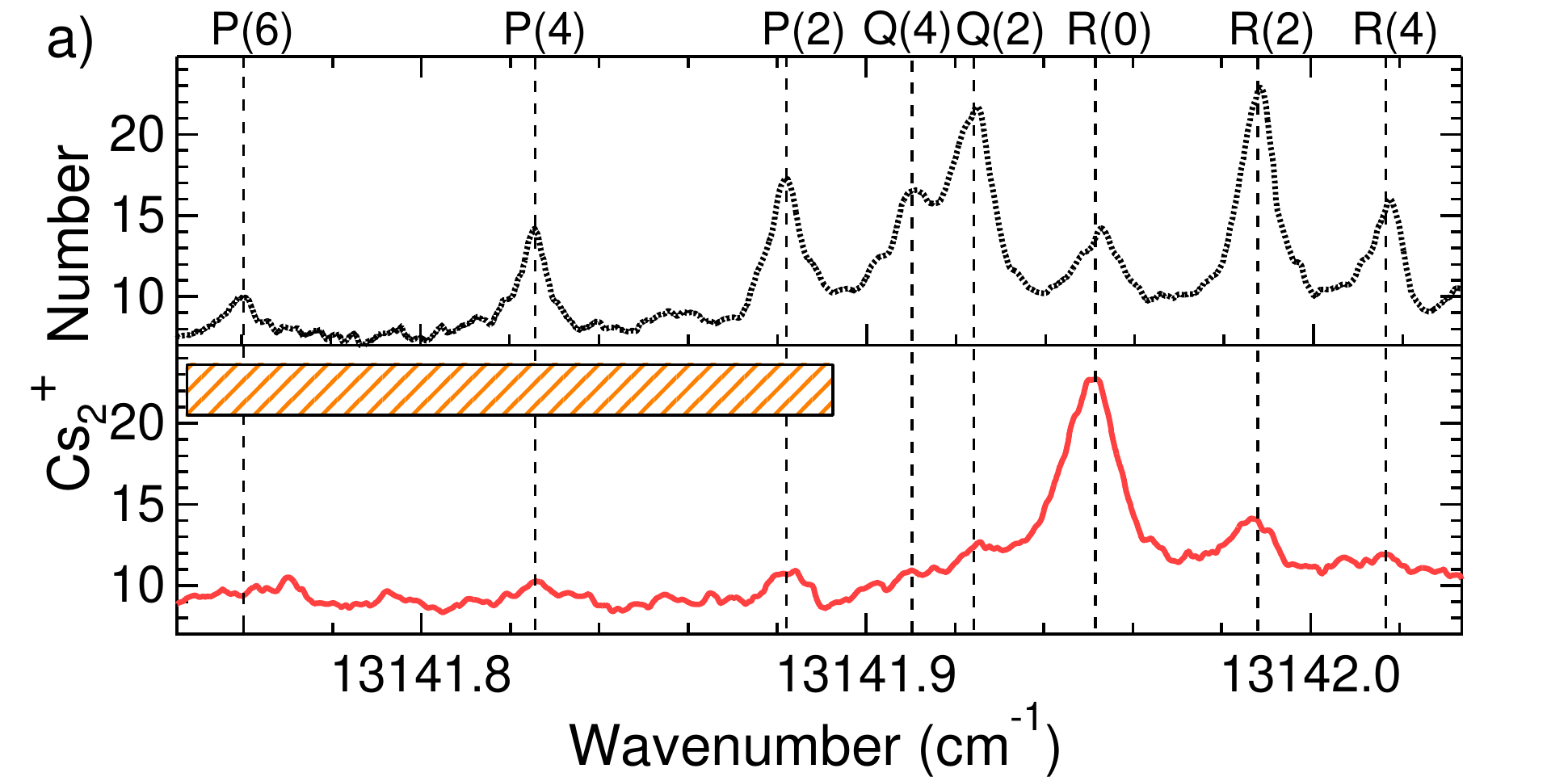}}}
\subfigure{
\resizebox*{7cm}{!}{\includegraphics{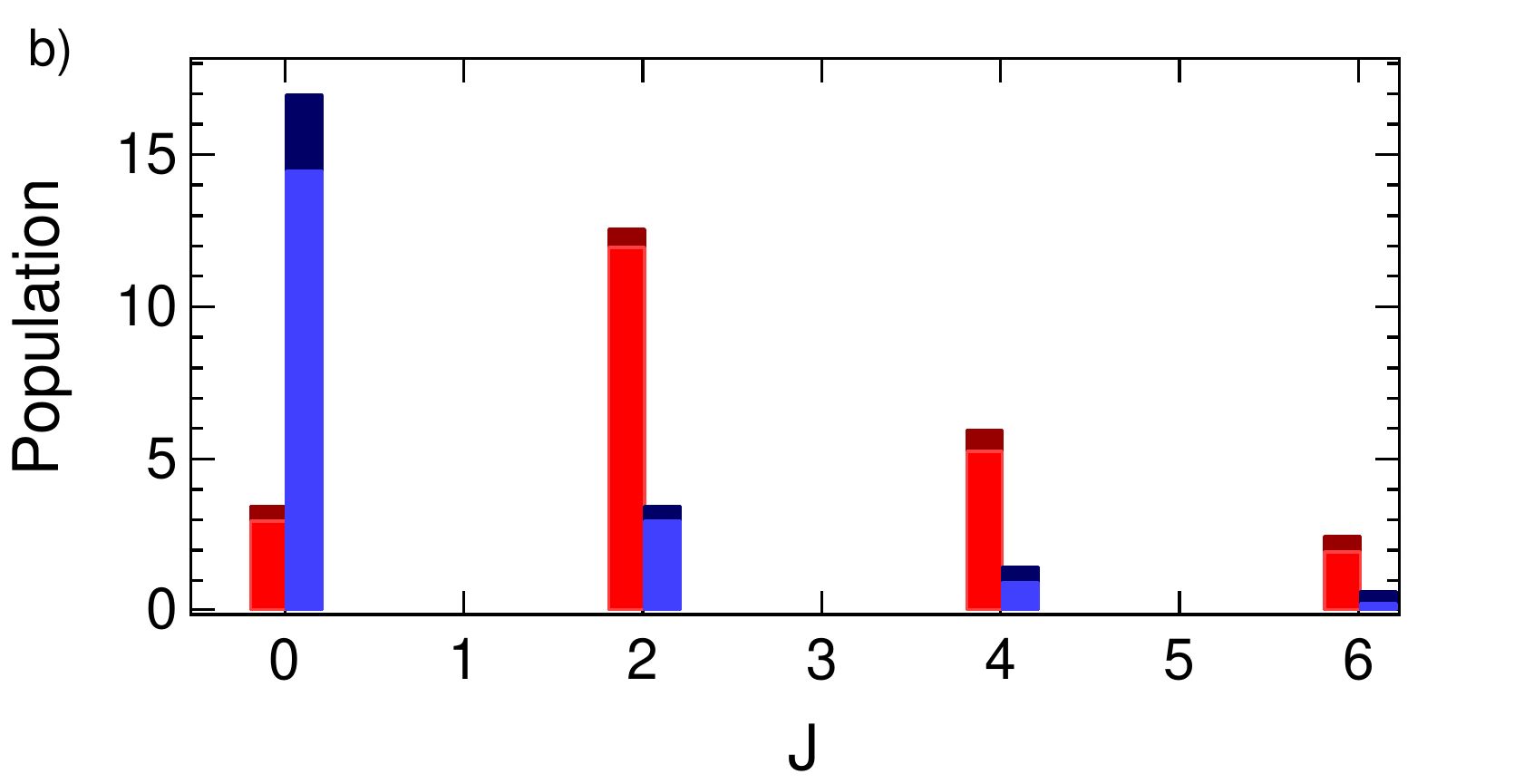}}}
\caption{a) Upper panel : Rotational spectrum, obtained by SpIDR, before rotational cooling is applied ; lower panel : rotational spectrum after rotational cooling is applied. The horizontal filled bar represents the frequency range scanned by the DFB rotational pumping laser. b) Population measured in different rotational levels, before (red bars) and after (blue bars) the cooling phase. The darker colors on the top of the bars indicate the error on the ion numbers.}
\label{sample-figure2}
\end{minipage}
\end{center}
\end{figure}

After a single absorption-spontaneous emission cycle, the rotational quantum number can decrease by two units or keep the same value, both with a similar probability, i.e. 1/2. However, as the vibrational level is modified by this operation, the vibrational pumping is required, and is likely to modify again the rotational quantum number. Figure \ref{Jchange_vibcooling} gives quantitative indications about the way vibrational cooling modifies the rotational quantum number. The ccumulation of a few cycles is thus enough to efficiently transfer populations to the absolute ro-vibrational ground state ($v=0, \rm J=0$). With the appropriate resolution, the hyperfine structure could in principle be resolved as well. 
We roughly estimate that the vibrational cooling takes about $100 \ \mu \rm{s}$ to bring back molecules to $v=0$ from any vibrational levels. In principle, below 100 $\mu \rm{s}$, a single ramp does not allow a complete rotational pumping, but several cycles give a pumping efficiency equivalent to that obtained with a 100 $\mu \rm{s}$ scan period. To estimate the vibrational cooling time, we have used the effective laser intensity on a resonance I$_{\rm laser}^{\rm effectif}$ = I$_{\rm laser} \frac{\Gamma}{\Gamma_{\rm laser}}$ where $\Gamma \approx (2 \pi)15$ MHz is the linewidth, I$_{\rm laser}$ the laser intensity, and $\Gamma_{\rm laser}$ the laser bandwidth ($\sim$ 100 cm$^{-1}$). The saturation intensity of a typical rovibrational transition is I$_{\rm sat} ^{\rm mol}=\frac{\rm I_{\rm sat}}{q_{{v', v"}} \ \rm HL}$, where I$_{\rm sat} \sim $ 2 mW/cm$^{2}$ is the saturation intensity of 2 levels system with lifetime $\tau$ = 15 ns, which is the molecular lifetime in excited state. We have considered typical Franck-Condon factors of q$_{v', v"}$ = 0.1 and Hönl-London factor of HL = $\frac{1}{4}$, I$_{\rm sat} ^{\rm mol}$ $\sim $ 160 mW/cm$^{2}$. I$_{\rm laser}^{\rm effectif}$ = 10$^{-4}$ mW/cm$^{2} \simeq \frac{\rm I_{\rm sat}^{\rm mol}}{1000}$ = $\frac{\rm I_{\rm sat}}{10^{5}}$. The minimum scan period ensuring a vibrational pumping is t = 10$^{5} \tau$ = 100 $\mu \rm{s}$.\\

In the experiment to favor a decrease of the $J$ quantum number, the rotational laser cooling is scanned over the B $^{1}\Pi_{u}, v_{B}=3, J-1 \leftarrow \rm{X} ^{1}\Sigma_{g}^{+}$, $v_{X}=0, \rm J = 2-6$, i.e. the laser spectrum only induces $J_{\rm B} = J_{\rm X}-1$ (P branch) transitions as shown in figure \ref{sample-figure2} a). In Figure \ref{sample-figure2} b), the rotational state distribution of 4 levels are presented both for the situation after vibrational pumping without rotational cooling and when the   two cooling lasers are applied. The real number of molecules in a specific rotational state is found by considering the q factor between $v_{\rm B}$ = 3 and  $v_{\rm X}$ = 7 and the microchannel plates efficiency.

\section{Conclusion}
We have demonstrated how to efficiently transfer an initial rovibrational distribution to a single rovibrational level. The method follows the same idea developed for the vibrational cooling \cite{MatthieuViteau07112008, 2009PhRvA..80e1401S, PhysRevLett.109.183001}, i.e. a broadband light source is designed to induce efficient absorption/spontaneous emission cycles that lead to a population redistribution getting closer to a predefined target. Adding the fact that we have previously demonstrated the possibility to use broadband light to transfer populations from a given electronic state to another one \cite{2012_PRA_Horchani_Conversion}, this work shows that the use of broadband source is really versatile and opens up the path to full (external plus internal degrees of freedom) laser cooling of molecules in cells or beams. \\
This article is dedicated to Bretislav Friedrich, whose work has given a great impetus to the difficult but nevertheless fascinating world of cold molecules.

\subsection{Acknowledgements}
Laboratoire Aimé Cotton is a member of Institut Francilien de Recherche sur les Atomes Froids (IFRAF) and of the LABEX PALM initiative. The exchange project between the University of Pisa and the University of Paris-Sud is acknowledged. A. F. and I. M. have been supported by the "Triangle de la Physique" under contracts No. 2007-n.74T and No. 2009-035T "GULFSTREAM" and No. 2010-097T-COCO2. The research leading to these results has received funding from the European Union Seventh Framework Programme FP7/2007-2013 under grant agreement n. 251391 COLDBEAMS and from the European Research
Council under the grant agreement n. 277762 COLDNANO. D.C. is the Coordinator or Principal Investigator of these projects. We thank O. Dulieu and N. Bouloufa-Maafa for fruitful discussions.


\end{document}